\documentclass[preprint,11pt,authoryear]{elsarticle}

\usepackage{endnotes}
\usepackage{graphicx}
\usepackage{booktabs}
\usepackage{subcaption}
\usepackage[top=3.2cm,bottom=2.2cm,left=2.5cm,right=2.5cm]{geometry}

\usepackage{threeparttable}
\usepackage{caption}
\usepackage{amssymb}
\usepackage{tabularx}
\usepackage{amsmath}
\usepackage{multirow}

\usepackage[dvipsnames,svgnames,table]{xcolor}
\usepackage[colorlinks,
linkcolor=red,
anchorcolor=blue,
citecolor=blue
]{hyperref}

\usepackage[utf8]{inputenc}
\begin{document}
\def\T{{\mathrm{\scriptscriptstyle \top}}}
\newcommand*{\dif}{\mathop{}\!\mathrm{d}}
\newcommand{\bA}{{\mathbf A}}
\newcommand{\bB}{{\mathbf B}}
\newcommand{\bH}{{\mathbf H}}
\newcommand{\bX}{{\mathbf X}}
\newcommand{\bY}{{\mathbf Y}}
\newcommand{\bZ}{{\mathbf Z}}
\newcommand{\bI}{{\mathbf I}}
\newcommand{\ba}{\boldsymbol {a}}
\newcommand{\bx}{\boldsymbol {x}}
\newcommand{\bu}{\boldsymbol {u}}
\newcommand{\by}{\boldsymbol {y}}
\newcommand{\bz}{\boldsymbol {z}}
\newcommand{\bv}{\boldsymbol {v}}
\newcommand{\be}{\boldsymbol {e}}
\newcommand{\bt}{\boldsymbol {t}}
\newcommand{\br}{{\mathbf r}}
\newcommand{\bU}{{\mathbf U}}
\newcommand{\bV}{{\mathbf V}}
\newcommand{\bW}{{\mathbf W}}
\newcommand{\bS}{{\mathbf S}}
\newcommand{\bR}{{\mathbf R}}
\newcommand{\bT} {\boldsymbol{T}}
\newcommand{\balpha} {\boldsymbol{\alpha}}
\newcommand{\bbeta}  {\boldsymbol{\beta}}
\newcommand{\bBeta}  {\boldsymbol{\Beta}}
\newcommand{\bdelta} {\boldsymbol{\delta}}
\newcommand{\blambda}{\boldsymbol{\lambda}}
\newcommand{\bepsilon}{\bolsymbol{\varepsilon}}
\newcommand{\bOmega}{\boldsymbol{\Omega}}
\newcommand{\bomega}{\boldsymbol{\omega}}
\newcommand{\bUpsilon}{\boldsymbol{\Upsilon}}
\newcommand{\bSigma}{\boldsymbol{\Sigma}}
\newcommand{\bDelta}{\boldsymbol{\Delta}}
\newcommand{\bgamma}{\boldsymbol{\gamma}}
\newcommand{\brho}{\mbox{\boldmath$\rho$}}
\newcommand{\bTheta} {\boldsymbol{\Theta}}
\newcommand{\bPhi} {\boldsymbol{\Phi}}
\newcommand{\bPsi} {\boldsymbol{\Psi}}
\newcommand{\btheta} {\boldsymbol{\theta}}
\newcommand{\bxi} {\boldsymbol{\xi}}
\newcommand{\bmu} {\boldsymbol{\mu}}
\newcommand{\bzeta} {\boldsymbol{\zeta}}
\newcommand{\bGamma} {\boldsymbol{\Gamma}}
\newcommand{\bLambda} {\boldsymbol{\Lambda}}
\newcommand{\bsigma}{\boldsymbol{\sigma}}
\newcommand{\bPi}{\boldsymbol{\Pi}}
\newcommand{\bzero}{{\mathbf 0}}
\newcommand{\bnu}{\boldsymbol{\nu}}
\newcommand{\bet}{\boldsymbol{\eta}}
\newcommand{\ve}{{\varepsilon}}
\newcommand{\cov}{{\rm Cov}}

\newcommand{\bvartheta}{\boldsymbol{\vartheta}}
\newcommand\code{\bgroup\@makeother\_\@makeother\~\@makeother\$\@codex}

\DeclareFontFamily{U}{FdSymbolC}{}

\newcommand{\e}{\mathbb{E}}
\newcommand{\EE}{\mathbb{E}}
\newcommand{\PP}{\mathbb{P}}
\newcommand{\RR}{\mathbb{R}}
\newcommand{\BB}{\mathbb{B}}

\newcommand{\argmin}{\mathop{\mathrm{argmin}}}
\newcommand{\argmax}{\mathop{\mathrm{argmax}}}

\newcommand{\cA}{\mathcal{A}}
\newcommand{\cB}{\mathcal{B}}
\newcommand{\cC}{\mathcal{C}}
\newcommand{\cD}{\mathcal{D}}
\newcommand{\cE}{\mathcal{E}}
\newcommand{\cF}{\mathcal{F}}
\newcommand{\cG}{\mathcal{G}}
\newcommand{\cH}{\mathcal{H}}
\newcommand{\cI}{\mathcal{I}}
\newcommand{\cJ}{\mathcal{J}}
\newcommand{\cK}{\mathcal{K}}
\newcommand{\cL}{\mathcal{L}}
\newcommand{\cM}{\mathcal{M}}
\newcommand{\cN}{\mathcal{N}}
\newcommand{\cO}{\mathcal{O}}
\newcommand{\cP}{\mathcal{P}}
\newcommand{\cQ}{\mathcal{Q}}
\newcommand{\cR}{\mathcal{R}}
\newcommand{\cS}{{\mathcal{S}}}
\newcommand{\cT}{{\mathcal{T}}}
\newcommand{\cU}{\mathcal{U}}
\newcommand{\cV}{\mathcal{V}}
\newcommand{\cW}{\mathcal{W}}
\newcommand{\cX}{\mathcal{X}}
\newcommand{\cY}{\mathcal{Y}}
\newcommand{\cZ}{\mathcal{Z}}
\begin{frontmatter}

\title{
Forecasting realized volatility in the stock market: a path-dependent perspective} 

\author[1]{Xiangdong Liu}
\ead{tliuxd@jnu.edu.cn}
\author[1]{Sicheng Fu}
\ead{wangxiaobo018@stu.jnu.edu.cn}

\author[2]{Shaopeng Hong\corref{cor1}}
\ead{124071400005@smail.swufe.edu.cn}
\cortext[cor1]{Corresponding author}

\affiliation[1]{organization={School of Economics, Jinan University},
            city={Guangzhou},
            postcode={510000}, 
            country={China}}

\affiliation[2]{organization={School of Statistics, Southwestern University of Finance and Economics},
            city={Chengdu},
            postcode={610074}, 
            country={China}}
\begin{abstract}
Financial market volatility prediction has been a research focus in recent years. This paper presents a new volatility prediction framework that captures volatility and trend features by leveraging long - and short - term memory in price data. It integrates the heterogeneous autoregressive (HAR) model family to develop a path-dependent heterogeneous autoregressive model family (HAR-PD). And, we propose a HAR-REQ model based on the empirical quantile as a threshold, which exhibits stronger forecasting ability compared to the HAR-REX model.
Subsequently, the predictive performance of the HAR-PD model family is evaluated by statistical tests using data from the Chinese stock market and compared with the benchmark HAR model family. The empirical results show that the HAR-PD model family has higher forecasting accuracy compared to the underlying HAR model family.
In addition, robustness tests confirm the significant predictive power of the HAR-PD model family.
\end{abstract}


\end{frontmatter}

\section{Introduction}
Volatility modeling has long been a key area of financial research. Traditionally, three types of models are considered: constant volatility models, local volatility models, and stochastic volatility models, e.g., \citep{black1973pricing, dupire1994pricing, heston1993closed}. 
However, constant volatility models and local volatility models have limitations in capturing the volatility smile phenomenon, making it difficult to model complex market dynamics effectively.

To address the limitations of traditional volatility models, \citet{foschi2008path} introduced an innovative path-dependent volatility (PDV) model. 
This model captures the dependence of volatility on the asset's historical price path by incorporating a deviation process, thereby providing a more accurate depiction of market behavior. 
Furthermore, \citet{doi:10.1080/07350015.2012.663261} underscored the significant advantages of the PDV model in forecasting volatility.

By leveraging the entire historical trajectory of the underlying asset prices, the PDV model can more comprehensively capture long-term memory effects and complex dynamics, and remain highly sensitive to new market environments. 
The PDV model maintains forecast accuracy at different time scales and has a high sensitivity to extreme market events, which is particularly important for predicting potential large fluctuations. \citet{parent2022rough} integrated the concept of path dependence with the rough Heston model \citep{euch2018perfect} to introduce the rough path-dependent volatility model. This model can capture a stronger Zumbach effect \endnote{The strong Zumbach effect refers to the phenomenon where the distribution of future realized volatility depends not only on past squared returns but also on the entire trajectory of past realized volatility. The weak Zumbach effect refers to the covariance between past squared returns and future realized volatility (within a given duration) being greater than the covariance between past realized volatility and future squared returns.}. These research findings not only enrich the theoretical foundation of PDV models but also open up new avenues for their practical application in finance.

\citet{guyon2023volatility} pointed out that the predictive power of the PDV model for volatility lies in its ability to replicate the dynamic characteristics of volatility and provide a reinterpretation based on trend and volatility features. On the other hand, considering the volatility clustering phenomenon, a series of forecasting models have been proposed, such as the ARCH, GARCH, and its extensions, see e.g., \citep{6ab571e5-c8f0-3fcd-9005-ed6f2adc76d7, glosten1993relation,haas2004new,creal2013generalized}. However, these models have difficulty dealing with long-term dependencies and rely mainly on low-frequency data, which limits their ability to capture the dynamics of high-frequency trading activities.

To address this issue, \citet{26b2223b-2816-35cd-8184-da686b05fa62} integrated realized volatility and constructed models such as the ARFIMA-RV suitable for high-frequency data, thereby improving volatility forecasting. \citet{10.1093/jjfinec/nbp001} pointed out the limitations of the model in capturing multi-scale dynamics. The HAR-RV model based on the heterogeneous market hypothesis solves this problem by integrating volatility components over different time horizons and significantly outperforms GARCH and ARFIMA-RV in financial volatility forecasting.
Other scholars have further studied realized volatility and discovered another important aspect of volatility dynamics - jumps in intraday returns. These sudden and significant changes are called jumps, and point out that realized volatility can be decomposed into continuous and jump components. \citet{lee2008jumps} emphasized the key role of jumps in volatility modeling and its impact on accurate risk assessment and prediction. \citet{lee2008jumps} highlighted the significant importance of jumps in volatility modeling. \citet{doi:10.1080/07350015.2012.663261} further proposed the HAR model with jumps. To delve deeper into the jump process, \citet{10.1162/REST_a_00503} assumed that the sign (positive or negative) of price jumps has different impacts on volatility. Using the realized semivariance estimator, \citet{barndorff2008measuring} decomposed realized volatility into positive semivariance and negative semivariance components.  \citet{clements2019moderate}, starting from the distribution characteristics of intraday returns, observed that during periods of financial turmoil, intraday returns may exhibit a heavy-tailed distribution, which affects volatility forecasting. Based on the tail distribution of returns, they defined extreme volatility and moderate volatility and proposed the HAR-REX model.

Synthesizing existing research discoveries, it has been demonstrated that high-frequency volatility models improved by methods such as jumps and realized semivariance (RS) can significantly enhance volatility forecasting accuracy. However, to date, no high-frequency volatility model has been constructed from the path-dependent perspective. 

Based on the above discussion, this paper puts forward a novel volatility forecasting framework. It brings together the PDV model concept and the HAR model family structure. 
This integration aims to systematically consider the recent trajectory of asset prices to more comprehensively and accurately capture the dynamic characteristics of financial asset volatility. 

The main contributions of our work are summarized as follows. 
\begin{enumerate}[(i)]
     \item In this research, we demonstrate that the Chinese stock market possesses notable path-dependent traits that are instrumental in explaining the underlying patterns of market volatility. Leveraging these insights, we have developed an innovative forecasting model designed to enhance the precision of stock market volatility predictions.
     \item Aiming to enhance predictive accuracy, we have refined the HAR-REX model by substituting the conventional cumulative normal distribution threshold with an innovative approach that employs empirical quantiles to re-segment realized volatility. Our empirical analyses reveal that the revamped HAR-REQ model, along with its extension, the HAR-PD-REQ model, exhibit significantly improved forecasting efficacy across daily, weekly, and monthly volatility horizons.
    \item Empirical results indicate that the HAR-PD model family, which is constructed based on path-dependent frameworks, exhibits superior predictive performance for forecasting realized volatility. The significance of the results persists even after undergoing various robustness tests, thereby confirming the reliability of the findings.
\end{enumerate}


The remainder of this paper is structured as follows. In Section \ref{sec:2}, we introduce the fundamental theories of volatility and the construction of volatility forecasting models. In Section \ref{sec:3}, we present all the benchmark models used in this study and the methods for combining models. In Section \ref{sec:4}, we discuss the volatility of the Chinese financial market from a path-dependent perspective and conduct an empirical analysis of the dataset. We use various statistical testing methods, including the Model Confidence Set (MCS) test \citep{hansen2011model} and the out-of-sample $R^2$ test \citep{campbell2008predicting}, to evaluate the predictive capabilities of the HAR model family and the HAR-PD model family, and we report and discuss the results. In Section \ref{sec:5}, we present the conclusion that the proposed HAR-PD model family demonstrates superior predictive ability.

\section{Theoretical framework}\label{sec:2}
Consider a complete filtered probability space $(\Omega, \mathcal{F}, \{\mathcal{F}_t\}_{t\geq 0}, \mathbb{P})$. The price $S_t$ of a financial asset is assumed to follow the stochastic differential equation (SDE):  
\begin{equation}\label{eq:j.d.model}
    \frac{dS_t}{S_t} = \mu_t \mathrm{d}t + \sigma_t \mathrm{d}W_t + \kappa_t \mathrm{d}q_t \,,
\end{equation}
where $\mu_t$ represents the drift term, $\sigma_t$ denotes the instantaneous volatility, $W_t$ is a standard Brownian motion, $q_t$ is a Poisson counting process, and $\kappa_t$ corresponds to the jump size.  

Quadratic variation is a pivotal concept in stochastic process theory, serving as a sophisticated measure to characterize the volatility and variational behavior of random process trajectories. The integrated volatility $\mathrm{IV_t}$ represents the squared instantaneous volatility integrated over a time interval $[0, T]$.
Integrated volatility is defined as:  
$$
\mathrm{IV}_t = \int_0^t \sigma_s^2 \,\mathrm{d}s\,,
$$
when $S_t$ does not contain jump components ($\kappa_t = 0$), the quadratic variation is equal to the integrated volatility, 
$$
    \mathrm{QV}_t = \mathrm{IV}_t = \int_0^t \sigma_s^2 \, \mathrm{d}s\,.
$$
However, when $X_t$ contains jump components ($\kappa_t \neq 0$), the quadratic variation is composed of both continuous and jump components, as shown by:
$$
\mathrm{QV}_t = 
\underbrace{\int_{0}^{t} \sigma_s^2 \, \mathrm{d}s}_{\text{continuous component}} + \underbrace{\sum_{0<s\leqslant t} \kappa_s^2}_{\text{jump component}} \neq \mathrm{IV}_t \,.
$$

\citet{26b2223b-2816-35cd-8184-da686b05fa62} defined RV as the sum of squared intraday returns from high-frequency data, which is expressed as:
\begin{equation} \label{rv}
\mathrm{RV}_{t} = \sum_{i=1}^n r_{t, i}^2 \xrightarrow{p} 
\int_0^t \sigma_s^2 \, \mathrm{d}s + \sum_{0<s \leqslant t} \kappa_s^2\,,
\end{equation}
where $r_{t, i} = \log S_{t,i} - \log S_{t,i-1}
$ is the $i$-th log return on day $t$. $S_{t,i}$ is the $i$-th log closing price on day $t$. 
They prove that under certain conditions, as the sampling frequency goes to infinity, realized volatility converges to the quadratic variation process in probability. 
\subsection{The PDV model}\label{secpd}

The PDV model assumes that the asset price $S_t$ follows a functional SDE \endnote{The well-posedness of \eqref{eq: FSDE} has been extensively studied and is beyond the scope of this paper; interested readers may refer to \citet{ning2021wellposedness} and \citet{cozma2018strong}}:
\begin{equation}\label{eq: FSDE}
\frac{dS_t}{S_t}=\mu_t dt+\sigma_t( {S_u(u\leq t)}, G_t)dW_t\,,
\end{equation}
where $\mu_t$ denotes a progressively measurable stochastic process, $W_t$ is a standard Brownian motion, and $S_u(u\leq t)$ 
is the trajectory of $S_u$ for $0$ to $t$, representing a functional mapping from the sample space $\RR$ to the path space $\mathcal{C}(\mathbb{R})$. 
In contrast to classical jump-diffusion model \eqref{eq:j.d.model}, the PDV models \eqref{eq: FSDE} postulate that the volatility is dependent upon the entire trajectory of asset $S_u(u\leq t)$ instead of $S_t$. 
This implies that the volatility is determined not only by the current price level but also by the path through which the price reaches its current level.
The path-dependent variable $G_t$ can incorporate various metrics, including the moving average ($S^\Delta_t$), exponentially weighted moments ($Y^\Delta_t$), and historical maximum or minimum prices ($Max^\Delta_t$ and $Min^\Delta_t$), where:
\begin{align*}
    S^\Delta_t &= \frac{\int_0^\Delta w_\tau S_{t-\tau} d\tau}{\int_0^\Delta w_\tau d\tau}\,,\quad Y^\Delta_t=\int_{-\infty}^\Delta\lambda e^{-\lambda(t-s)}\log\big\{S_t/S_s \cdot e^{-r(t-s)}\big\} ds ,\\
    Max^\Delta_t &= \min_{t-\tau\leq u \leq t}S_u\,,\quad 
    Min^\Delta_t = \min_{t-\tau\leq u \leq t}S_u \,.
\end{align*}
Here, $\Delta$ represents the time window length. For a one-day window, $\Delta$ is set to $78$, corresponding to the number of five-minute intervals in a typical 6.5-hour trading day ($\Delta = 6.5 \times 60 / 5$). 
The term $w_\tau$ in $S^{\Delta}_t$ represents the weight function for time interval $\tau$, where $w_\tau = 1$ corresponds to constant weights, $w_\tau = \tau$ to linear weights, and $w_\tau = e^{-\lambda \tau}$ to exponential weights, with $\lambda > 0$ denoting the decay factor.

As demonstrated above, the PDV models establish that asset volatility is determined not only by current market conditions but also by the historical trajectory of asset prices. In other words, volatility dynamics are contingent upon past price movements, particularly the influence of historical volatility on current volatility levels. The fundamental premise of the model posits that market participants' behavior and market sentiment are significantly influenced by historical events. Through these inherent characteristics, the PDV models effectively capture volatility clustering, leverage effects, and asymmetric time-reversal properties via their explicit incorporation of historical price information, thereby enabling more accurate volatility predictions.
To further elucidate the path-dependent properties of asset volatility, two key features have been developed: the trend feature ${R}_{1,t}$ and the volatility feature ${R}_{2,t}$. The trend feature, ${R}_{1,t}$, characterizes the historical movement of asset prices, thereby reflecting long-term market trends, while the volatility feature measures historical price volatility levels, capturing short-term market fluctuations. Through the incorporation of these two features, the model comprehensively captures the dynamic evolution of asset volatility while accounting for market participants' reactions to historical events. The trend feature is constructed as a weighted sum of past returns, designed to capture the trend features of asset prices, expressed as:
\begin{equation}\label{(8)}
{R}_{1, t}:=\sum_{i\textless t}K_{\lambda_1}(t-i)\tilde{r}_{t-i}\,,\quad 
    \tilde{r}_{t-i}=\frac{S_{t}-S_{t-i}}{S_{t-i}}\,,
\end{equation}
where $K_{\lambda_1}$ is a kernel function that assigns different weights to past daily returns based on the lag term $t-i$. 
Moreover, 
$K_{\lambda_1}$  is specified as exponential kernels:
$$
    {{K}_{\lambda}(\tau)=\lambda e^{-\lambda \cdot \tau}, \lambda > 0}\,,
$$
where the parameters $\lambda$ determine the degree of influence past returns exert on current volatility.

\begin{figure}
    \centering
    \includegraphics[width=0.8\linewidth]{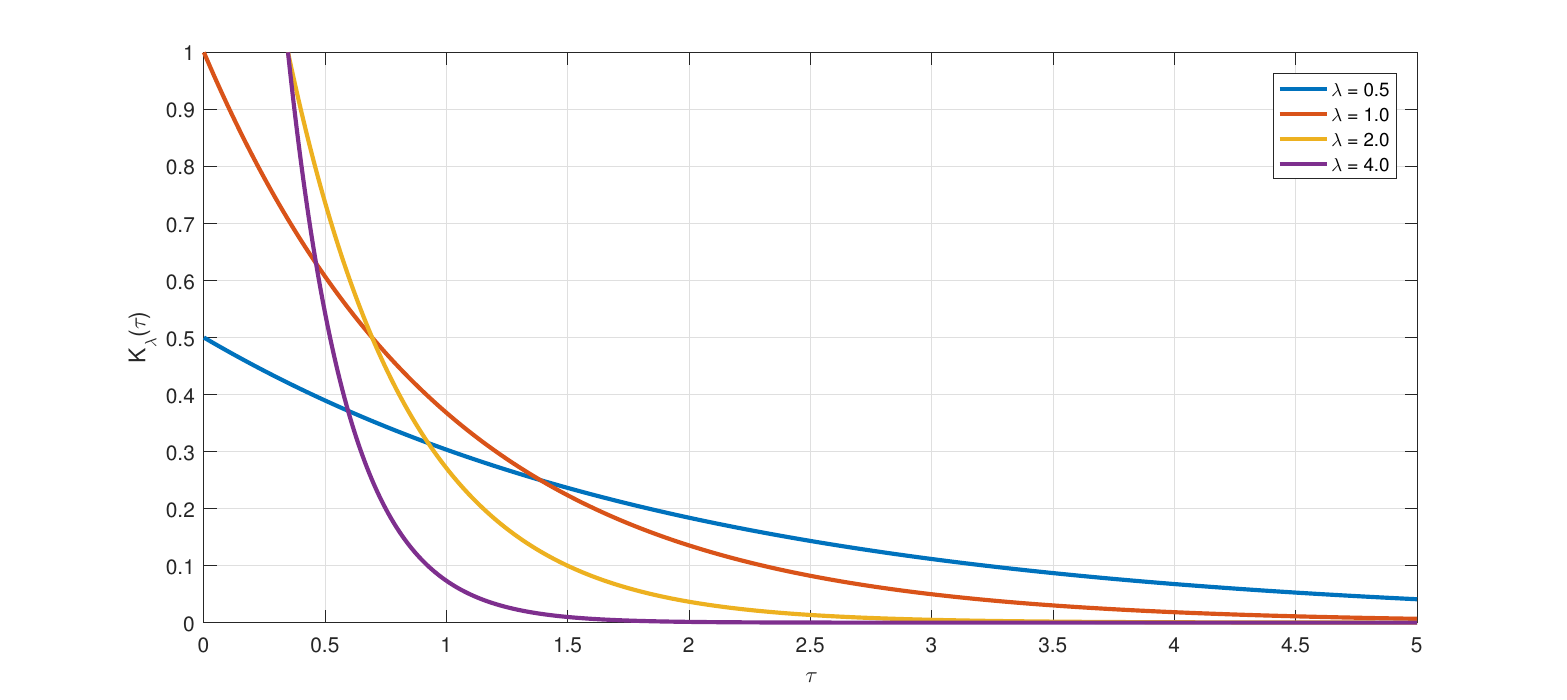}
    \caption{Exponential kernel function}
    \label{fig:tau}
\end{figure}

As shown in Figure \ref{fig:tau}, the exponential kernel function decays differently across various $\lambda$ values with respect to the time interval $\tau$. When $\lambda$ = 4, the decay is most rapid, causing the model to prioritize recent returns while diminishing the influence of historical ones. Larger $\lambda$ values reduce path dependency and reflect short-term dynamics, which is useful in fast-changing markets. In contrast, smaller $\lambda$ values slow decay, allowing historical returns to significantly influence volatility and capturing long-memory effects, making them more suitable for markets with strong path dependency.
Furthermore, the volatility characteristic, defined as the sum of squared past returns, is employed to capture recent price volatility and is expressed as:
\begin{equation}\label{r2t}
{R}_{2, t}:=\sum_{i\textless t}{K}_{\lambda_2}(t-i) \tilde{r}_{t-i}^2\,.
\end{equation}

\cite{guyon2023volatility} demonstrated that the PDV models quantify volatility over time by incorporating trend and volatility features, thereby capturing both long-term and short-term effects. The fundamental model is expressed as:
\begin{equation}\label{eqrv}
\mathrm{RV}_{t}=\beta_0+\beta_1{R}_{1, t}+\beta_2{{R}_{2, t}}+\epsilon_t\,, \end{equation}
where $\beta_0$ represents a constant term, $\beta_1\leq0$ denotes the sensitivity parameter for the trend variable ${R}_{1,t}$, and $\beta_2\geq0$ serves as the sensitivity parameter for ${R}_{2, t}$, which measures recent market price activity.

\subsection{Realized volatility decomposition}
\subsubsection{Continuous and discrete components}
\citet{barndorff2004power} introduced the realized bipower variation ($\mathrm{RBV}_t$), and under certain conditions, they proved that $\mathrm{RBV}_t$ is a consistent estimator of the integrated volatility. The expression for $\mathrm{RBV}_t$ is given as follows:
$$
  \mathrm{RBV}_{t} =  \frac{n}{\mu_{1}^2(n-1)}  \sum_{i=2}^{n} \left|r_{t, i-1}\right| \left|r_{t, i}\right|\,,
$$
where $\mu_{1} = \sqrt{2/\pi}$, and the factor $n/(n-1)$ is used to adjust for sample size. \citet{barndorff2003realized} demonstrated that under the assumption that the price process $S_t$ is a semimartingale, $\mathrm{RBV}_t$ serves as a consistent estimator of the integrated volatility and the difference between ${\rm RV}_t$ and $\mathrm{RBV}_t$ provides an estimator for jump variation. 
$$
\mathrm{RBV}_t \xrightarrow{p} \int_0^t \sigma_s^2 \, \mathrm{d}s\,,~~~{\rm and}~~~\mathrm{RV}_t - \mathrm{RBV}_t \xrightarrow{p} \sum_{0<s \leqslant t} \kappa_s^2\,.
$$
Moreover, \citet{huang2005relative} proposed a $Z$-statistic for jump identification and to remove the influence of these insignificant jumps. 
The $Z$-statistic is defined as:
$$
   Z_t = \frac{\sqrt{n} (\mathrm{RV}_t - \mathrm{RBV}_t) \mathrm{RV}_t^{-1}}{\sqrt{\left(\mu_{1}^{-4} + 2\mu_{1}^{-2} - 5 \right) \max \left(1, \frac{\mathrm{RTQ}_t}{\mathrm{RBV}_t^2}\right)}}\,,
$$
where $\mathrm{RTQ}_t$ is the realized tripower quarticity, given by:
$$
   \mathrm{RTQ}_t =\frac{n^2}{ \mu_{4/3}^3(n-4)}  \sum_{k=5}^{n} \left|r_{t, k-4}\right|^{4/3}\cdot \left|r_{t, k-2}\right|^{4/3}\cdot \left|r_{t, k}\right|^{4/3}\,,
$$ 
where $\mu_{4/3} = 2^{2/3}\Gamma(1/6)/(6\pi^{1/2})$. 
 Based on the properties of this indicator function, the significant jumps can be isolated:
$$
\mathrm{CJ}_{t, \alpha} = I(Z_t > \phi_{1-\alpha}) (\mathrm{RV}_t - \mathrm{RBV}_t)\,,
$$ 
where $I(\cdot)$ is indicator function, $\phi_{1-\alpha}$ is the $(1-\alpha)$-quantile of the standard normal distribution. Here $\alpha$ is set to 0.05 and is used in the rest of the paper.

\subsubsection{Realized semivariance}
\citet{barndorff2008measuring} posited that, in financial markets, downside risk is typically more significant than upside risk for most investors, as it directly relates to potential losses. Based on this observation, the realized semivariance is introduced as a novel risk measure, focusing on capturing price fluctuations during periods of negative returns. Realized semivariance distinguishes between positive and negative intraday logarithmic returns and calculates the sum of squared positive intraday logarithmic returns to derive both positive and negative semivariance. The expression is as follows:
\begin{equation*}
\begin{aligned}
\mathrm{RS}_t^-=\sum_{i=1}^nr_{t, i}^2{{I(r_{t, i}<0)}}\,,\quad 
\mathrm{RS}_t^+=\sum_{i=1}^nr_{t, i}^2{{I(r_{t, i}>0)}}\,.
\end{aligned}
\end{equation*}

\subsubsection{Extreme and moderate volatility}\label{sec:empirical quantile}
Extreme events in financial markets, such as financial crises, large-scale market crashes, and circuit breakers, can significantly impact volatility. These events often lead to dramatic changes in market volatility over a short period, which traditional volatility measures may fail to capture effectively. Furthermore, market volatility exhibits dynamic multiscale characteristics, meaning both short-term and long-term volatility components coexist. Traditional volatility measures and forecasting models may not adequately capture this multiscale nature, particularly the long-term component's impact. To address these issues, \citet{clements2019moderate} proposed a new decomposition method, which divides realized volatility into moderate and extreme realized volatility. The procedure is as follows:
\begin{enumerate}[(i)]
    \item \textbf{Threshold determination:}For daily time $t$, calculate the unconditional volatility $\sigma_t$ of the daily return, where $t = 1, 2, \dots, N$, $N$ is the number of days. Then choose the tail probability $\alpha$, and define the positive and negative thresholds according to this probability as:
    \[
    r_t^- = F^{-1}(\alpha)\sigma_t \,,\quad r_t^+ = F^{-1}(1-\alpha)\sigma_t\,,
    \]
    where $F^{-1}$ is the inverse cumulative distribution function of the normal distribution.

    \item \textbf{Moderate Realized Volatility Construction:} The moderate realized volatility ($\mathrm{REX}_t^m$) is calculated by squaring the returns for each period and summing those that lie between the positive and negative thresholds. Specifically, a threshold is set for each day $t$ and if $r_t^- \leq r_{t,i} \leq r_t^+$, the squared return $r_{t,i}^2$ is added to the moderate volatility construction:
    \[
    \mathrm{REX}_t^m = \sum_{i=1}^n r_{t,i}^2 {I}(r_t^- < r_{t,i} < r_t^+)\,.
    \]
    \item \textbf{Extreme Volatility Construction:} Extreme volatility is divided into positive extreme realized volatility, $\mathrm{REX}_t^+$, and negative extreme realized volatility, $\mathrm{REX}_t^-$. A threshold is set for each day $t$ and if $r_{t,i} \geq r_t^+$, the squared return $r_{t,i}^2$ is added to the positive extreme volatility construction; if $r_{t,i} \leq r_t^-$, it is added to the negative extreme volatility construction:
    \[
    \mathrm{REX}_t^- = \sum_{i=1}^n r_{t,i}^2 {I}(r_{t,i} \leq r_t^-)\,, \quad \mathrm{REX}_t^+ = \sum_{i=1}^n r_{t,i}^2 {I}(r_{t,i} \geq r_t^+)\,.
    \]
\end{enumerate}

Using this method, volatility is decomposed into three components: moderate volatility ($\mathrm{REX}_t^m$), which captures the long-term component; positive extreme volatility ($\mathrm{REX}_t^+$), and negative extreme volatility ($\mathrm{REX}_t^-$), which together capture the short-term component.

The positive and negative thresholds are defined by selecting the tail probability $\alpha$. Subsequently, the positive and negative thresholds are set as $r_t^- = F^{-1}(\alpha)\sigma_t$ and $r_t^+ = F^{-1}(1-\alpha)\sigma_t$, based on which extreme and moderate volatility are computed. \cite{clements2019moderate} sets $F^{-1}$ as the CDF of the standard normal distribution, but
financial returns often exhibit the characteristics of leptokurtosis, heavy-tailed distributions, or asymmetry.
We suggest using empirical quantile to replace $F^{-1}(\alpha)\sigma_t$. The 0.95 quantile is taken as the positive threshold, with the portion of intraday returns exceeding this threshold designated as extreme positive volatility. Similarly, the 0.05 quantile is taken as the negative threshold, with the portion of intraday returns below this threshold designated as extreme negative volatility. The portion between the positive and negative thresholds is classified as moderate volatility. The expressions are given as follows:
\begin{equation*}
\begin{aligned}
\mathrm{REXQ}_t^- &=\sum_{i=1}^nr_{t,i}^2I(r_{t,i}\leq \hat{Q}_{t,0.05}),\\
\mathrm{REXQ}_t^+ &=\sum_{i=1}^nr_{t,i}^2I(r_{t,i}\geq \hat{Q}_{t,0.95}),\\
\mathrm{REXQ}_t^m &=\sum_{i=1}^nr_{t,i}^2I(\hat{Q}_{t,0.05}<r_{t,i}<\hat{Q}_{t,0.95})\,.
\end{aligned}
\end{equation*}
Where $\hat{Q}_{t,0.05}$ is based on the actual returns within each trading day $t$ to determine the thresholds for extreme and moderate volatility, thereby providing a basis for the decomposition of volatility. The construction HAR-REQ is applied in \eqref{harreq}.

\section{Volatility prediction models}
\label{sec:3}
\subsection{Benchmark: HAR model family}\label{benchmark}
Drawing upon the heterogeneous market hypothesis, \citet{10.1093/jjfinec/nbp001} developed a model that integrates realized volatility across multiple time scales to capture both long-term and short-term volatility dynamics in financial markets, expressed as:
\begin{equation}\label{eq:HAR}
\mathrm{RV}_t=\beta_0+\beta_1\mathrm{{RV}}_{t-1}+\beta_2\mathrm{\overline{RV}}_{t-5}+\beta_3\mathrm{\overline{RV}}_{t-22}+\epsilon_t\,.
\end{equation}
The model comprises three explanatory variables: $\mathrm{{RV}}_{t-1}$, $\mathrm{\overline{RV}}_{t-5}$, and $\mathrm{\overline{RV}}_{t-22}$, which correspond to short-term (daily), medium-term (weekly), and long-term (monthly) volatility, respectively. The coefficients $\beta_0, \beta_1$, $\beta_2$, $\beta_3$ in \eqref{eq:HAR} can be estimated using the loss function MSE for estimation. The lagged terms $\mathrm{\overline{RV}}_{t-5}$ and $\mathrm{\overline{RV}}_{t-22}$ represent five-step and twenty-two-step lags, respectively, and can be generalized as $\mathrm{\overline{RV}}_{t-h}$, where $h$ denotes the number of lag terms, calculated as:
\begin{equation}\label{meanlag}
    \mathrm{\overline{RV}}_{t-h}={\frac1h(\mathrm{RV}_{t-1}+\cdots+\mathrm{RV}_{t-h})}\,.
\end{equation}

\citet{andersen2007roughing} and \citet{corsi2009har} developed a new HAR model through the decomposition of continuous and jump volatility components. The model is constructed as follows:
$$
\mathrm{RV}_t=\beta_0+\beta_1\mathrm{{CV}}_{t-1}+\beta_2\mathrm{\overline{CV}}_{t-5}+\beta_3\mathrm{\overline{CV}}_{t-22}+\beta_4\mathrm{{CJ}}_{t-1}+\beta_5\mathrm{\overline{CJ}}_{t-5}+\beta_6\mathrm{\overline{CJ}}_{t-22}+\epsilon_t\,,
$$
where $\mathrm{{CV}}_{t-1}$ and $\mathrm{{CJ}}_{t-1}$ denote the first-order lags of $\mathrm{{CV}}_t$ and $\mathrm{{CJ}}_t$, respectively. The terms $\mathrm{\overline{CV}}_{t-5}$, $\mathrm{\overline{CV}}_{t-22}$, $\mathrm{\overline{CJ}}_{t-5}$, and $\mathrm{\overline{CJ}}_{t-22}$ are constructed similarly to \eqref{meanlag}.

Drawing upon realized semivariance, \cite{10.1162/REST_a_00503} introduced the HAR-RS model, which incorporates both positive and negative semivariance components:
$$
\mathrm{RV}_{t}=\beta_{0}+\beta_{1}\mathrm{RS}_{t-1}^{+}+\beta_{2}\mathrm{\overline{RS}}_{t-5}^{+}+\beta_{3}\mathrm{\overline{RS}}_{t-22}^{+}+\beta_{4}\mathrm{RS}_{t-1}^{-}+\beta_{5}\mathrm{\overline{RS}}_{t-5}^{-}+\beta_{6}\mathrm{\overline{RS}}_{t-22}^{-}+\epsilon_{t}\,.
$$

\citet{clements2019moderate} proposed the HAR-REX model by incorporating extreme volatility and bidirectional moderate volatility components:
\begin{equation}\label{harrex}
\begin{split}
    \mathrm{RV}_{t}=\beta_{0}+\beta_{1}\mathrm{REX}_{t-1}^{+}+\beta_{2}\mathrm{\overline{REX}}_{t-5}^{+}+\beta_{3}\mathrm{\overline{REX}}_{t-22}^{+}+\beta_{4}\mathrm{REX}_{t-1}^{-}+\beta_{5}\mathrm{\overline{REX}}_{t-5}^{-}
+\\ \quad \beta_{6}\mathrm{\overline{REX}}_{t-22}^{-}+\beta_{7}\mathrm{REX}_{t-1}^{m}+\beta_{8}\mathrm{\overline{REX}}_{t-5}^{m}+\beta_{9}\mathrm{\overline{REX}}_{t-22}^{m}+\epsilon_{t}\,,
\end{split}
\end{equation}
where the construction of variables in \eqref{harrex} is similar to that in \eqref{meanlag}.

The HAR-REQ model is constructed using empirical quantiles and can be expressed as follows:
\begin{equation}\label{harreq}
\begin{split}
\mathrm{RV}_{t}=&\beta_{0}+\beta_{1}\mathrm{REQ}_{t-1}^{+}+\beta_{2}\mathrm{\overline{REQ}}_{t-5}^{+}+\beta_{3}\mathrm{\overline{REQ}}_{t-22}^{+}+\beta_{4}\mathrm{REQ}_{t-1}^{-}+\beta_{5}\mathrm{\overline{REQ}}_{t-5}^{-}+
\\&\quad\beta_{6}\mathrm{\overline{REQ}}_{t-22}^{-}+\beta_{7}\mathrm{REQ}_{t-1}^{m}+\beta_{8}\mathrm{\overline{REQ}}_{t-5}^{m} +\beta_{9}\mathrm{\overline{REQ}}_{t-22}^{m}+\epsilon_{t}\,,
\end{split}
\end{equation}
where $\mathrm{REQ}_{t-1}^-$, $\mathrm{\overline{REQ}}_{t-5}^-$, and $\mathrm{\overline{REQ}}_{t-22}^-$ represent the negative extreme volatility based on daily, weekly, and monthly empirical quantiles, respectively. The other components are denoted in a similar manner.

\subsection{Encoding path-dependent: HAR-PD models}
The structure of \eqref{eqrv} from Section \ref{secpd} is incorporated into the HAR model family to construct the HAR-PD model family. 
As established in Section \ref{secpd}, it is clear that $R_{2,t}$ as a volatility feature is more explanatory of realized volatility, along with the structure of the trend feature $R_{1,t}$. The HAR-RV model is restructured by incorporating path dependence to form the HAR-PD-RV model:
\begin{equation}\label{har-pd-rv}
    \mathrm{RV}_{t}=\beta_{0}+\beta_{1}{R}_{2, t-1}+\beta_{2}{\overline{R}}_{2, t-5}+\beta_{3}{\overline{R}}_{2, t-22}+\epsilon_{t}\,,
\end{equation}

where ${R}_{2, t-1}$, ${\overline{R}}_{2, t-5}$, and ${\overline{R}}_{2, t-22}$ represent volatility features based on the moments ${t-1}$, ${t-5}$, and ${t-22}$, respectively. The construction method for ${\overline{R}}_{2,t-h}$ is similar \eqref{r2t}.

Considering the decomposition of jump and continuous components, the HAR-PD-CJ model is formulated as follows:
\begin{equation}\label{harcjpd}
\begin{split}
\mathrm{RV}_{{t}} &= \beta_{0}+\beta_{1}{R}_{2, t-1}+\beta_{2}{\overline{R}}_{2, t-5}+\beta_{3}{\overline{R}}_{2, t-22} +\beta_{4}\mathrm{{PDCJ}}_{t-1}+\beta_{5}\mathrm{\overline{{PDCJ}}}_{t-5}+ \\
&\quad \beta_{6}\mathrm{\overline{{PDCJ}}}_{t-22}+\beta_{7}\mathrm{{PDCV}}_{t-1}+\beta_{8}\mathrm{\overline{{PDCV}}}_{t-5}+\beta_{9}\mathrm{\overline{{PDCV}}}_{t-22}+\epsilon_{t}\,.
\end{split}
\end{equation}
The construction method of $\mathrm{\overline{PDCJ}}_{t-h}$ can be found in \eqref{meanlag}.

In the HAR-PD-CJ model, the volatility characteristics incorporate both jump and continuous components. The volatility characteristics describe the overall fluctuation within a specific time period, while jumps in financial markets indicate significant changes in asset prices. By incorporating jump characteristics, volatility predictions can be improved, particularly for fluctuations caused by public events, announcements, and other market uncertainties. Here, $\mathrm{{PDCJ}}_{t-1}$, $\mathrm{\overline{{PDCJ}}}_{t-5}$, and $\mathrm{\overline{{PDCJ}}}_{t-22}$ represent the jump components based on the path dependence at time $t-1$, $t-5$, $t-22$ respectively. Combined with the constructor of \eqref{(8)}, the $\mathrm{PDCJ}_t$ can be constructed as:
\begin{equation}\label{jisuan}
    \mathrm{{PDCJ}}_{t}=
\sum_{i\textless t}K_{\lambda}(t-i)\mathrm{CJ}_{t-i}\,.
\end{equation}

Considering the decomposition of positive and negative variance, the HAR-PD-RS model is formulated as follows:
\begin{equation*}
\begin{aligned}
\mathrm{RV}_{t}=&\beta_{0}+\beta_{1}{R}_{1, t-1}+\beta_{2}{\overline{R}}_{1, t-5}+\beta_{3}{\overline{R}}_{1, t-22}+\beta_{4}\mathrm{{PDRS}}_{t-1}^{+}+\beta_{5}\mathrm{\overline{{PDRS}}}_{t-5}^{+}+\\
&\quad \beta_{6}\mathrm{\overline{{PDRS}}}_{t-22}^{+}+\beta_{7}\mathrm{{PDRS}}_{t-1}^{-}+\beta_{8}\mathrm{\overline{{PDRS}}}_{t-5}^{-}+\beta_{9}\mathrm{\overline{{PDRS}}}_{t-22}^{-}+\epsilon_{t}\nonumber\,.
\end{aligned}
\end{equation*}

The benchmark HAR-RS model decomposes the realized semivariance into positive and negative components for volatility prediction. This decomposition captures both the actual price skewness within specific time periods and the dynamic evolution of realized skewness, thereby providing insights into volatility patterns. The construction method for $\mathrm{PDRS}_{t}$ is shown in \eqref{jisuan}, and the construction methods for $\mathrm{\overline{{PDRS}}}_{t-h}^{+}$ and $\mathrm{\overline{{PDRS}}}_{t-h}^{-}$ are shown in \eqref{meanlag}.

The HAR-PD-REX model, based on the cumulative distribution function of the normal distribution and involving the decomposition of volatility into extreme and moderate components, is formulated as follows:
\begin{align*}
\mathrm{RV}_{t} &= \beta_{0} + \beta_{1}{R}_{1, t-1} + \beta_{2}{\overline{R}}_{1, t-5} + \beta_{3}{\overline{R}}_{1, t-22} + \beta_{4}\mathrm{{PDREX}}_{t-1}^{+} + \beta_{5}\mathrm{\overline{{PDREX}}}_{t-5}^{+} + \\ & \quad \beta_{6}\mathrm{\overline{{PDREX}}}_{t-22}^{+}  
+ \beta_{7}\mathrm{{PDREX}}_{t-1}^{-} + \beta_{8}\mathrm{\overline{{PDREX}}}_{t-5}^{-} + \beta_{9}\mathrm{\overline{{PDREX}}}_{t-22}^{-} + \\ & \quad \beta_{10}\mathrm{{PDREX}}_{t-1}^{m} + \beta_{11}\mathrm{\overline{{PDREX}}}_{t-5}^{m} + \beta_{12}\mathrm{\overline{{PDREX}}}_{t-22}^{m} + \epsilon_{t} \,,
\end{align*}
where, the construction methods of $\mathrm{{PDREX}}_{t-1}^{+}$, $\mathrm{{PDREX}}_{t-1}^{-}$, and $\mathrm{{PDREX}}_{t-1}^{m}$ are similar to \eqref{jisuan}.

The HAR-PD-REQ model is constructed based on the method of empirical quantiles as follows:
\begin{align*}
\mathrm{RV}_{t}=&\beta_{0}+\beta_{1}{R}_{1, t-1}+\beta_{2}{\overline{R}}_{1, t-5}+\beta_{3}{\overline{R}}_{1, t-22}+\beta_{4}\mathrm{{PDREQ}}_{t-1}^{+}+\beta_{5}\mathrm{\overline{{PDREQ}}}_{t-5}^{+}+ \\ & \quad   \beta_{6}\mathrm{\overline{{PDREQ}}}_{t-22}^{+}+\beta_{7}\mathrm{{PDREQ}}_{t-1}^{-}
+\beta_{8}\mathrm{\overline{{PDREQ}}}_{t-5}^{-}+\beta_{9}\mathrm{\overline{{PDREQ}}}_{t-22}^{-}
+\\ & \quad \beta_{10}\mathrm{{PDREQ}}_{t-1}^{m}+\beta_{11}\mathrm{\overline{{PDREQ}}}_{t-5}^{m}+\beta_{12}\mathrm{\overline{{PDREQ}}}_{t-22}^{m}+\epsilon_{t}\,.
\end{align*}
The construction methods of $\mathrm{{PDREQ}}_{t-1}^{+}$, $\mathrm{\overline{{PDREQ}}}_{t-5}^{+}$, and $\mathrm{\overline{{PDREQ}}}_{t-22}^{+}$ are similar to those of $\eqref{meanlag}$ and $\eqref{jisuan}$.
\subsubsection{Shrinkage: LASSO-HAR-PD models}\label{sec:lasso}
The introduction of path-dependent features, while enhancing the predictive capability of models, may also lead to an increase in the number of model parameters, thereby causing overfitting. To demonstrate that the enhancement of predictive performance is not solely due to an increase in the number of variables, several strategies have been proposed by researchers. For instance, \citet{ding2021forecasting} and \citet{audrino2016lassoing}  suggested incorporating LASSO \citep{tibshirani1996regression} into the model to control the number of parameters, thereby effectively preventing overfitting. Additionally, \citet{burnham2004multimodel} proposed that the complexity of models could be assessed by incorporating information criteria (such as AIC and BIC), and these criteria could be used to compare the predictive accuracy of models, thereby further elucidating the balance between model complexity and predictive performance.  Write $\bbeta = (\beta_1,\ldots,\beta_d)^\T$. Suppose  the sample size is $N$,  the loss function of LASSO-HAR-PD models can be written as:
$$
\mathcal{L}(\bbeta)=\frac{1}{2}\sum_{t=1}^N\left(RV_t-
\bX_t^\T\bbeta\right)^2+\lambda\sum_{j=1}^p|\beta_j|\,.
$$
where $\lambda$ is  the tuning parameters we selecting by $10$-fold validation, and $\bX_t$ is a vector that comprises the covariates of interest. 
For instance, in \eqref{harcjpd}, 
$\bX_t = (R_{2,t-1},{\overline{R}}_{2, t-5},{\overline{R}}_{2, t-22},\mathrm{\overline{{PDCJ}}}_{t-5},\mathrm{\overline{{PDCV}}}_{t-5})^\T$.

\section{Empirical analysis}\label{sec:4}

\subsection{The Path-dependency property of
realized volatility}\label{sub3}
In order to argue that the Chinese stock market is also characterized by path-dependence, we compare a path-dependence model that includes the historical price paths of stocks with a non-path-dependence model that assumes that price movements are independent of past paths. The data used are from the Wind database. 
Considering the non-negligible microstructure noise in the commodity market \citep{ait2005often,zhang2005tale}, five-minute high-frequency data are selected as suggested by \citet{corsi2008volatility}. The datasets \endnote{The data that support the findings of this study are available from the
corresponding author upon reasonable request.} include the SSE (000001), CSI 300 (000300), SSE 50 (000016), GEI (399006), SSE 50 ETF (510050), and STAR 50 (000688). 
The SSE covers the period from January 4, 2005, to December 29, 2023. The SSE 50 covers the same period. The CSI 300 covers the period from April 8, 2005, to December 29, 2023. The GEI covers the period from June 1, 2010, to December 29, 2023. The SSE 50 ETF covers the period from February 23, 2005, to December 29, 2023. The STAR 50 covers the period from July 23, 2020, to December 29, 2023. 
The realized volatility model based on the path dependence property is constructed as follows:
\begin{equation}\label{hanyoupd}
    \mathrm{RV}_{t}=\beta_0+\beta_1{R}_{1, t}+\beta_2{{R}_{2, t}}+\epsilon_t\,.
\end{equation}

Consider a volatility model that lacks path-dependent properties, specifically one without an exponential kernel function:
\begin{equation}\label{nopd}
    \mathrm{RV}_{t,null} = \beta_0 + \beta_1 r_{t} + \beta_2 r_{t}^2 + \epsilon_t\,.
\end{equation}
In this context, \eqref{nopd} denotes the method for calculating returns based on non-path-dependence, where $r_t$ is defined by \eqref{(8)}. Meanwhile, the power kernel function ${K}_{\lambda}(\tau)$ is chosen to compute \eqref{hanyoupd}.

\begin{table}[!htb]
\centering
\caption{Results of parameter estimation with and without path-dependence for each index}
\tabcolsep 9pt
\label{tab:tabletable1}
\begin{threeparttable}
\begin{tabular}{@{}lccccccl@{}}
\toprule
 & \multicolumn{5}{c}{Non-path-dependent parameter estimates} \\
          \midrule
& $\lambda_1$ & $\lambda_2$ & $\beta_0$ & $\beta_1$ & $\beta_2$ & $Adj. R^2$ \\
\midrule
$\mathrm{RV}_{null}^{001}$ &  &  & $0.000^{**}$ & $-0.001^{***}$ & $0.385^{***}$ & 0.352 \\
&  &  & (0.000) & (0.000) & (0.008) &  \\
$\mathrm{RV}_{null}^{50}$ &  &  & $0.000^{***}$ & $-0.000^{*}$ & $0.378^{***}$ & 0.343 \\
& &  & (0.000) & (0.000) & (0.007) &  \\
$\mathrm{RV}_{null}^{300}$ & &  & $0.000^{***}$ & $-0.001^{***}$ & $0.367^{***}$ & 0.334 \\
& & & (0.000) & (0.000) & (0.007) &  \\
$\mathrm{RV}_{null}^{006}$ &  &  & $0.000^{***}$ & $-0.002^{***}$ & $0.312^{***}$ & 0.245 \\
&  &  & (0.000) & (0.000) & (0.010) &  \\
$\mathrm{RV}_{null}^{050}$ &  &  & $0.000^{***}$ & $-0.000$ & $0.397^{***}$ & 0.129 \\
&  &  & (0.000) & (0.000) & (0.015) &  \\
$\mathrm{RV}_{null}^{688}$ &  &  & $0.000^{***}$ & $-0.000$ & $0.145^{***}$ & 0.129 \\
&  &  & (0.000) & (0.000) & (0.012) &  \\
\hline\hline
          & \multicolumn{5}{c}{Path-dependent parameter estimates} \\
                \midrule
               & $\lambda_1$ & $\lambda_2$ & $\beta_0$ & $\beta_1$ & $\beta_2$ & $Adj. R^2$ \\
\midrule 
$\mathrm{RV}^{001}$ & $0.534^{***}$ & $0.520^{***}$ & $0.000^{***}$ & $-0.005^{***}$ & $0.778^{***}$ & 0.581 \\
& $(0.000)$ & $(0.060)$ & $(0.000)$ & $(0.032)$ & $(0.125)$ & \\
$\mathrm{RV}^{50}$ & $0.432^{***}$ & $0.408$ & $0.00^{***}$ & $-0.002^{***}$ & $0.825^{***}$ & 0.573 \\
& (0.000) & (0.173) & (0.000) & (0.000) & (0.010) &  \\
$\mathrm{RV}^{300}$ & $0.731^{***}$ & $0.326^{***}$ & $0.000$ & $-0.003^{***}$ & $0.852^{***}$ & 0.565 \\
& (0.000) & (0.057) & (0.000) & (0.000) & (0.015) &  \\
$\mathrm{RV}^{006}$ & $0.838^{***}$ & $0.453^{***}$ & $0.000$ & $-0.006^{**}$ & $0.742^{***}$ & 0.518 \\
& (0.000) & (0.003) & (0.000) & (0.000) & (0.098) &  \\
$\mathrm{RV}^{050}$ & $0.809^{***}$ & $0.644^{***}$ & $0.000$ & $-0.002^{*}$ & $0.790^{***}$ & 0.221 \\
& (0.000) & (0.005) & (0.000) & (0.001) & (0.022) &  \\
$\mathrm{RV}^{688}$ & $0.571^{***}$ & $0.649^{***}$ & $0.000^{*}$ & $-0.001^{*}$ & $0.310^{***}$ & 0.235 \\
& (0.001) & (0.005) & (0.000) & (0.000) & (0.019) &  \\
\bottomrule
\end{tabular}
\begin{tablenotes} 
\item{Note: $Adj. R^{2}$ denotes the adjusted goodness-of-fit. $*$, $**$, $***$ denote the rejection of the original hypothesis at 10$\%$, 5$\%$ and 1$\%$ significance levels, standard deviation of each parameter is shown in parentheses, and all estimation results are retained to three decimal places.}
\end{tablenotes}
\end{threeparttable} 
\end{table}

\begin{figure}[!htbp]
\centering
\includegraphics[scale=0.5]{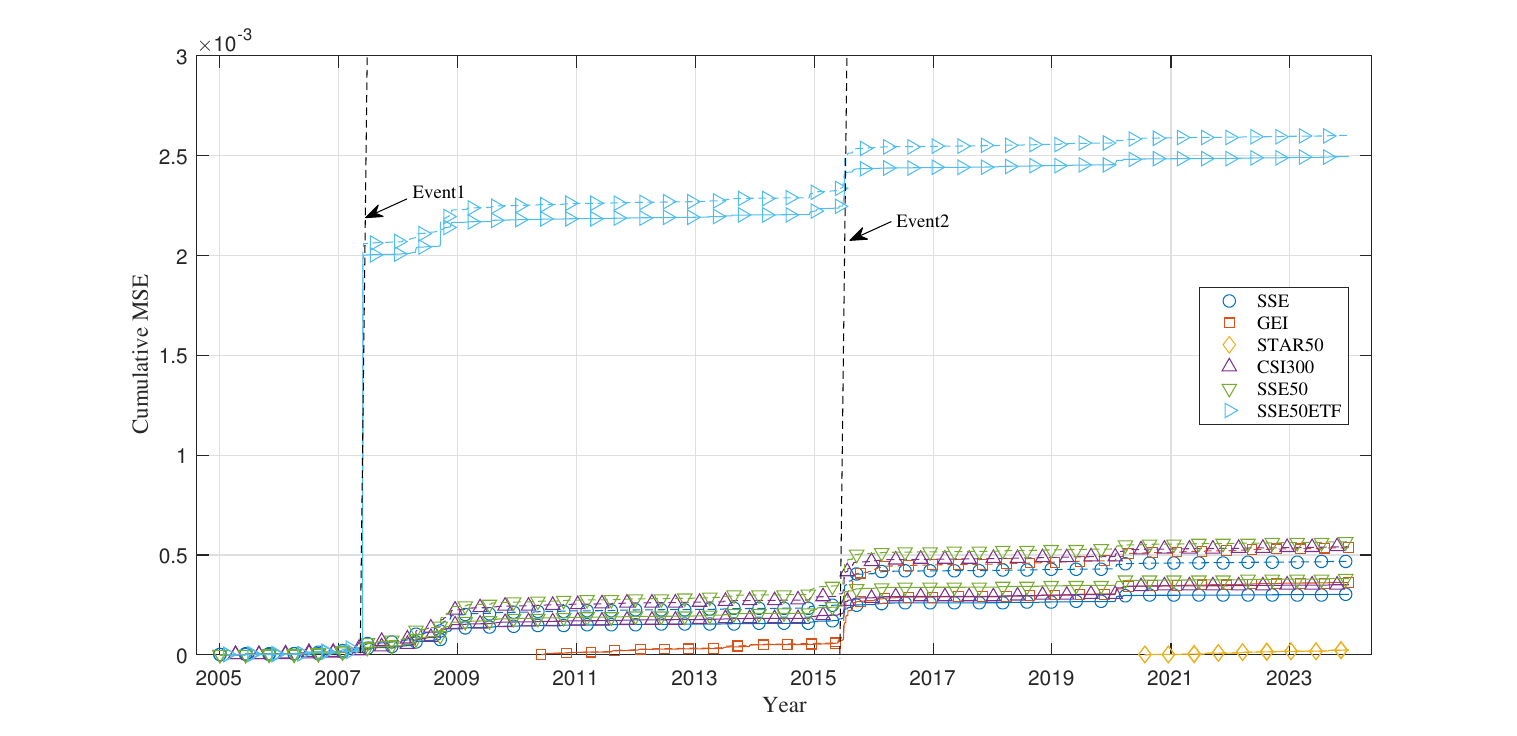}
\caption{Cumulative mean square error of realized volatility and path-dependent characteristics }
\label{figl}
 
\end{figure}

Table \ref{tab:tabletable1} reports the parameter estimates of the path-dependent and non-path-dependent models under high-frequency data for the Chinese stock market. The trend feature $(R_{1,t})$ and volatility feature $(R_{2,t})$ used in the path-dependent model are constructed based on the historical price paths of assets. The parameter estimation results show that the adjusted $R^2$ of the path-dependent model is much higher than the adjusted $R^2$ of the non-path-dependent model under all data, compared to the non-path-dependent model, which assumes that price changes are independent of past paths. Among them, the lowest goodness of fit is \(\mathrm{RV}_{{null}}^{050}\) and the highest is \(\mathrm{RV}^{001}\).

In order to visualize the path-dependent characteristics of the Chinese stock market, Figure \ref{figl} shows the cumulative mean-square error of realized volatility \eqref{hanyoupd} constructed based on the path-dependent characteristics and the realized volatility \eqref{nopd} of the non-path-dependent characteristics (without index kernel function) in the Chinese stock market in comparison, respectively. The data used include SSE (000001), CSI 300 (000300), SSE 50 (000016), GEI (399006), SSE 50 ETF (510050), and STAR 50 (000688). Where the horizontal axis indicates time and the vertical axis indicates mean square error. All solid lines denote the cumulative mean-squared error of the realized volatility constructed by path dependence, and all dashed lines denote the cumulative mean-squared error of the realized volatility constructed by non-path dependence. The black dashed line in the figure indicates the time period when the cumulative mean-squared error of each stock shows abnormal fluctuations.
\begin{enumerate}[Event 1:]
    \item At that time, the Chinese stock market experienced an unprecedented 'bull market'. This volatility was primarily driven by strong economic growth and increased global capital flows, leading to a sharp rise in investor optimism, which pushed market valuations to extremely high levels. This optimism reflected market participants' positive expectations for future economic prospects and a general confidence in sustained growth potential.
    \item [Event 2:] From early 2015 to mid-June, the Chinese stock market rose rapidly, with the Shanghai Composite Index and Shenzhen Component Index achieving gains beyond predictions, mainly driven by retail investors who were significantly influenced by leveraged funds. However, as a market bubble formed and regulatory policies were implemented, the stock market began to plummet sharply after mid-June, resulting in significant market volatility. In response to the rapid market decline, the government implemented a series of intervention measures, including suspending IPOs, implementing a circuit breaker mechanism, and deploying the 'national team' to stabilize the market.
\end{enumerate}

By analyzing Events 1 and 2, we find that the model with path-dependent characteristics is able to accurately fit the market volatility. 
As shown in Figure \ref{figl}, the path-dependent model utilizing past price path information exhibits a smaller prediction error than the model with non-path-dependent characteristics, and this difference is statistically significant. 
Overall, the empirical results confirm the validity of the path-dependent model in the Chinese stock market, with a high degree of fit between the model and the actual volatility.

\begin{figure}[htb]
    \centering
    \begin{minipage}{0.5\textwidth}
        \centering
        \includegraphics[width=\linewidth]{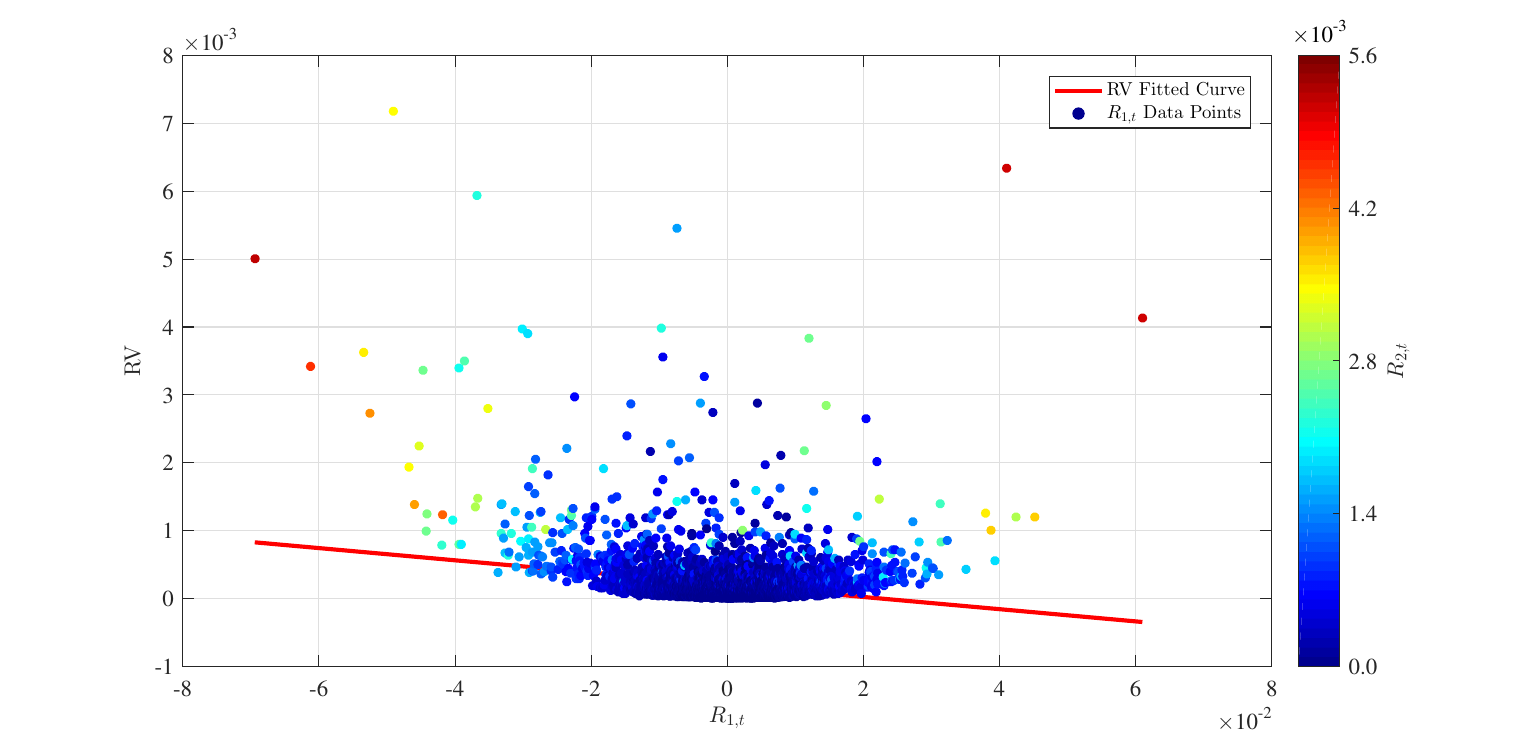} 
        \caption{Realized volatility and trend features}
        \label{fig:figure10}
    \end{minipage}\hfill
    \begin{minipage}{0.5\textwidth}
        \centering
        \includegraphics[width=\linewidth]{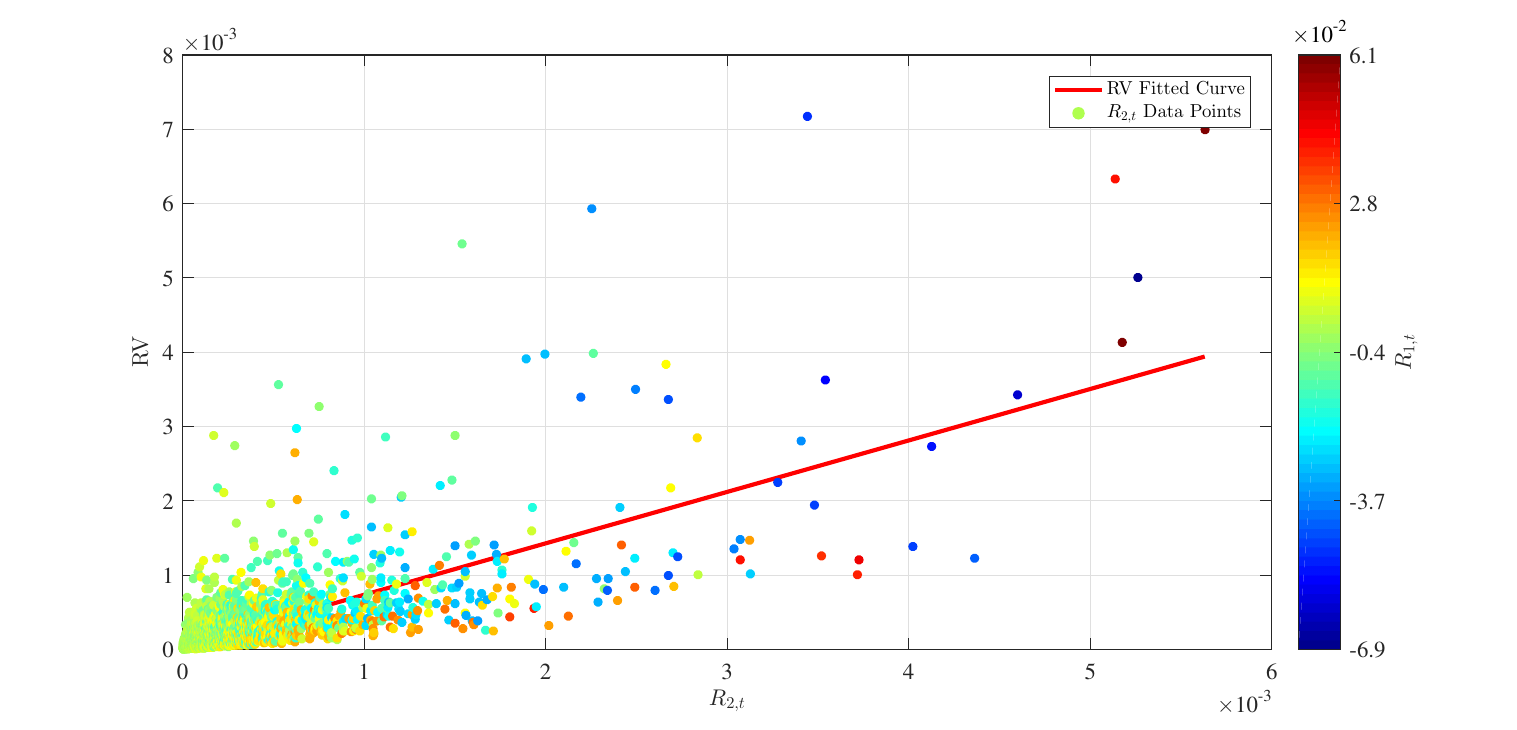}
        \caption{Realized volatility and volatility features}
        \label{fig:figure11}
    \end{minipage}
\end{figure}

As shown in Figures \ref{fig:figure10} and \ref{fig:figure11}, the relationship between the realized volatility of the SSE  and the trend feature (${R}_{1,t}$) as well as the volatility feature (${R}_{2,t}$) is illustrated. The red solid line represents the regression results of $\mathrm{RV}_t$ on ${R}_{1,t}$ and ${R}_{2,t}$, which are computed as follows:
$$
\mathrm{RV}_t= \beta_0 + \beta_1{R}_{1,t}+ \epsilon_t\,,~~\text{and}~~\mathrm{RV}_t= \beta_0 + \beta_1{R}_{2,t}+ \epsilon_t\,.
$$

As shown in Figure \ref{fig:figure10}, the scatter represents the value of $R_{1,t}$ and the color of the scatter indicates $R_{2,t}$. As ${R}_{1,t}$ increases to positive values, real volatility decreases rapidly and then stabilizes, and larger values of $\mathrm{RV}_t$ (outliers) appear to correspond to larger values of $R_{2,t}$ (warmer colors). On the other hand, as shown in Figure \ref{fig:figure11}, the scatter represents ${R}_{1,t}$, with the color changing from blue to red as ${R}_{1,t}$ increases. As a weighted sum of squared returns, ${R}_{2,t}$ is directly related to market volatility, with realized volatility being a linear function of volatility trend. For a given ${R}_{2,t}$, lower values of ${R}_{1,t}$ are usually associated with higher realized volatility.

\subsection{Contribution of Path-Dependence to Volatility}\label{pd}
To comprehensively evaluate the effectiveness of the path-dependent model in volatility forecasting, different models are introduced and compared to explore the explanatory power of path-dependent models with varying complexity. Specifically, explanatory power is demonstrated by progressively incorporating different path-dependent characteristics. Based on the construction method of path-dependent proposed by \citet{parent2023investigating}, the path-dependent can be formulated in the following basic forms:

\begin{align}
   M_1:\mathrm{RV}_t &= \beta_0 + \beta_1 R_{1,t} + \beta_3 \sqrt{R_{2,t}} \label{1}\,,\\
   M_2: \mathrm{RV}_t &= \beta_0 + \beta_1 R_{1,t} + \beta_3 R_{2,t} \label{2}\,, \\
   M_3:\mathrm{RV}_t &= \beta_0 + \beta_1 R_{1,t} \label{3}\,,\\
   M_4:\mathrm{RV}_t &= \beta_0 + \beta_1 \lvert R_{1,t} - \overline{R}_{1,t} \rvert \label{4}\,, \\
   M_5:\mathrm{RV}_t &= \beta_0 + \beta_1 R^2_{1,t}\,. \label{5}
\end{align}

The distinction between model $M_1$ and model $M_2$  lies in the nonlinear construction of volatility characteristics. Model $M_3$ serves as the benchmark model, assessing the explanatory power of a single trend feature on volatility. Model $M_4$ captures the dispersion of volatility by incorporating the absolute deviation of the trend feature from its mean, reflecting volatility explanatory power in extreme market conditions. Model $M_5$ emphasizes the direct contribution of the trend feature to volatility.

\begin{figure}
    \centering
    \begin{subfigure}{0.3\textwidth}
        \centering
        \includegraphics[width=\textwidth]{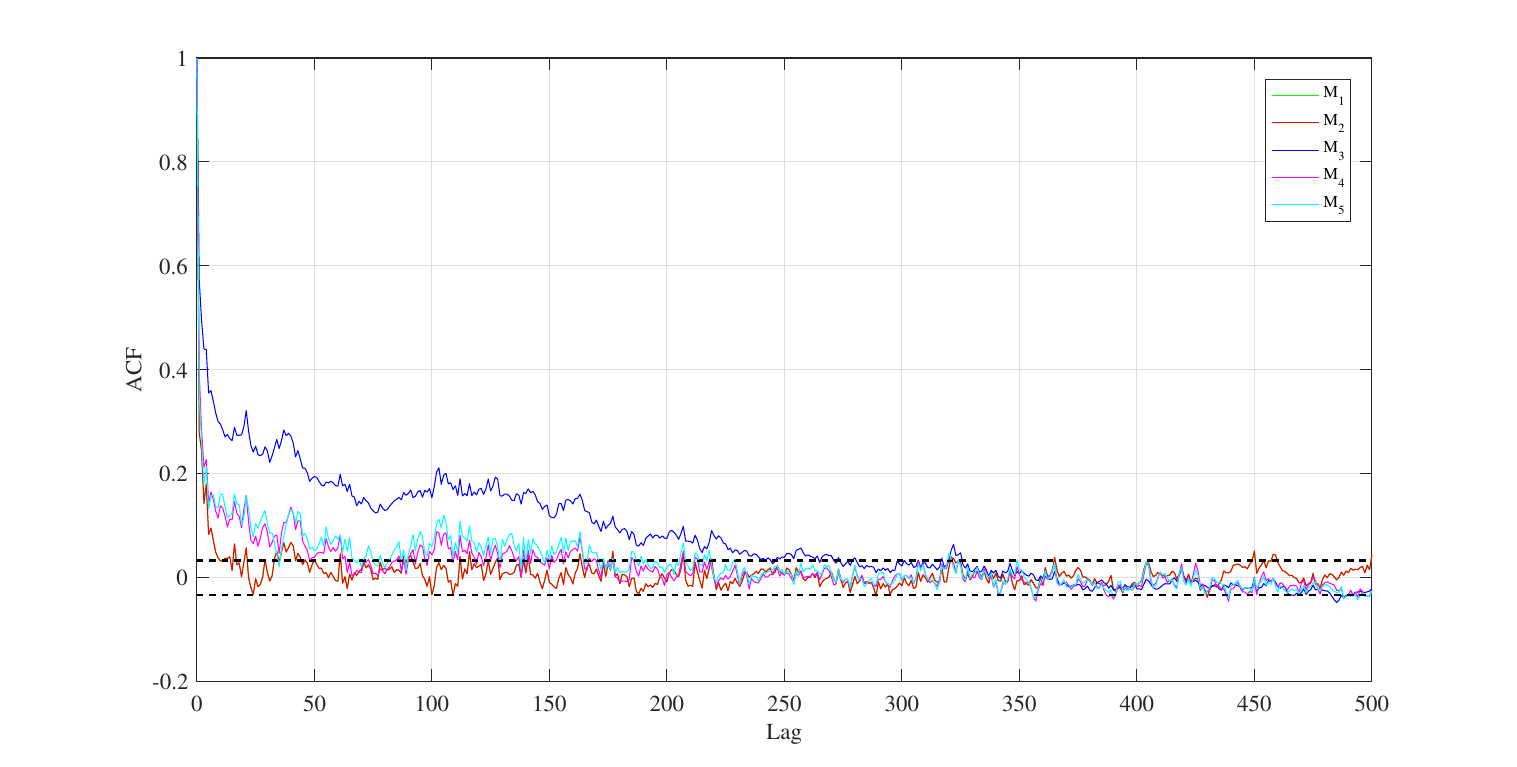}
        \caption{SSE residuals ACF under models $M_1$--$M_5$}
        \label{fig:sub1}
    \end{subfigure}
    \begin{subfigure}{0.3\textwidth}
        \centering
        \includegraphics[width=\textwidth]{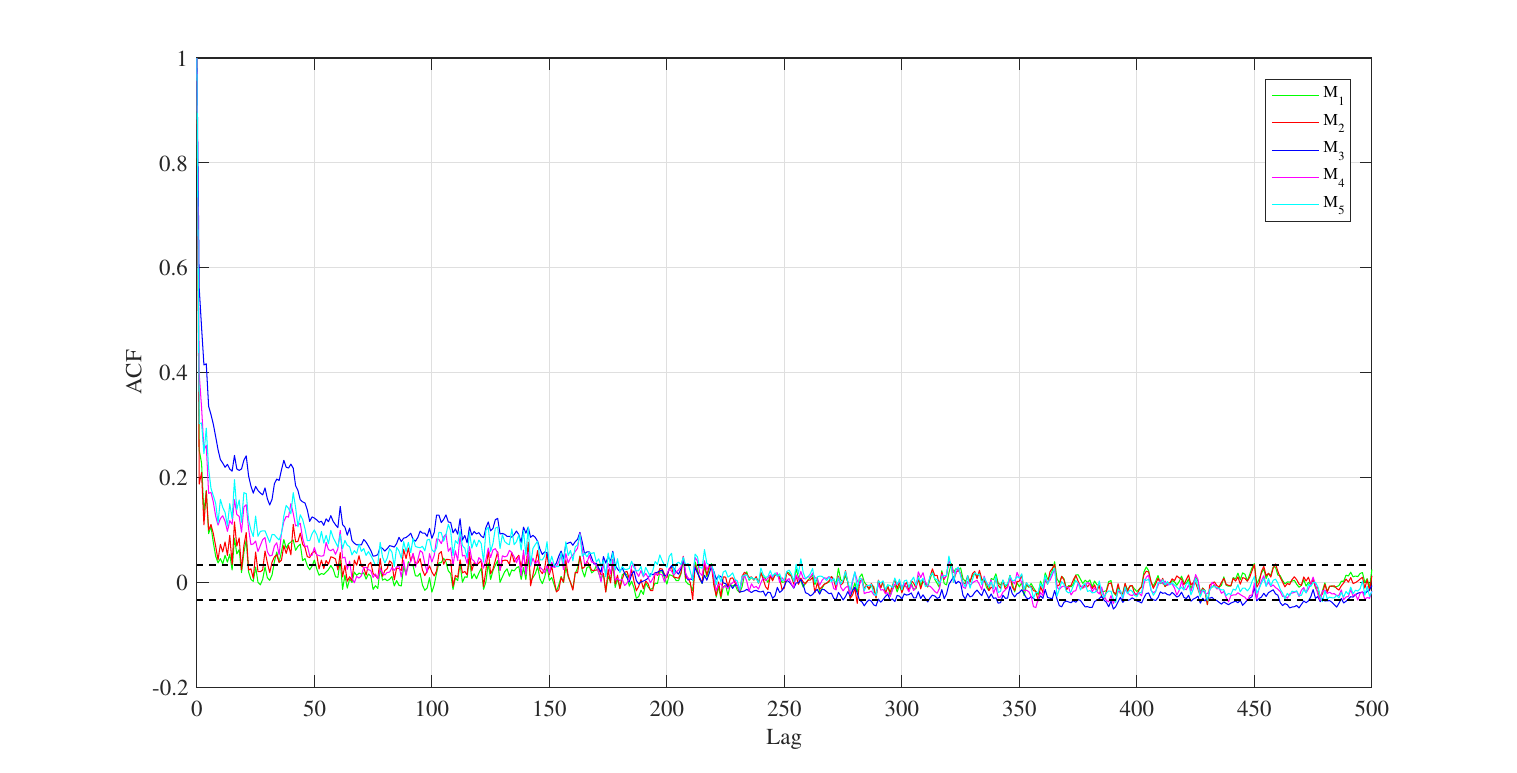}
        \caption{CSI 300 residuals ACF under models $M_1$--$M_5$}
        \label{fig:sub2}
    \end{subfigure}
    \begin{subfigure}{0.3\textwidth}
        \centering
        \includegraphics[width=\textwidth]{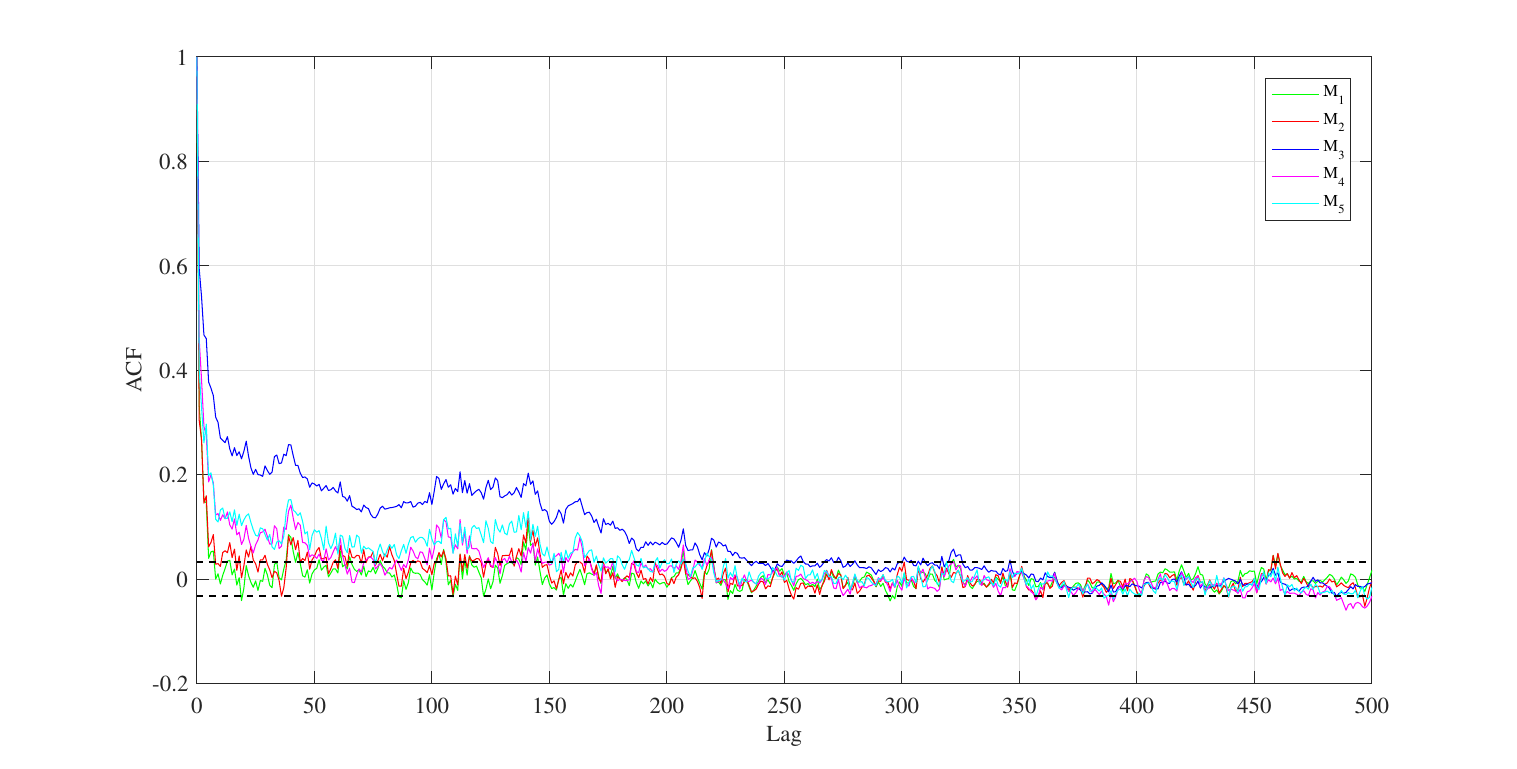}
        \caption{SSE 50 residuals ACF under models $M_1$--$M_5$}
        \label{fig:sub3}
    \end{subfigure}

    \begin{subfigure}{0.3\textwidth}
        \centering
        \includegraphics[width=\textwidth]{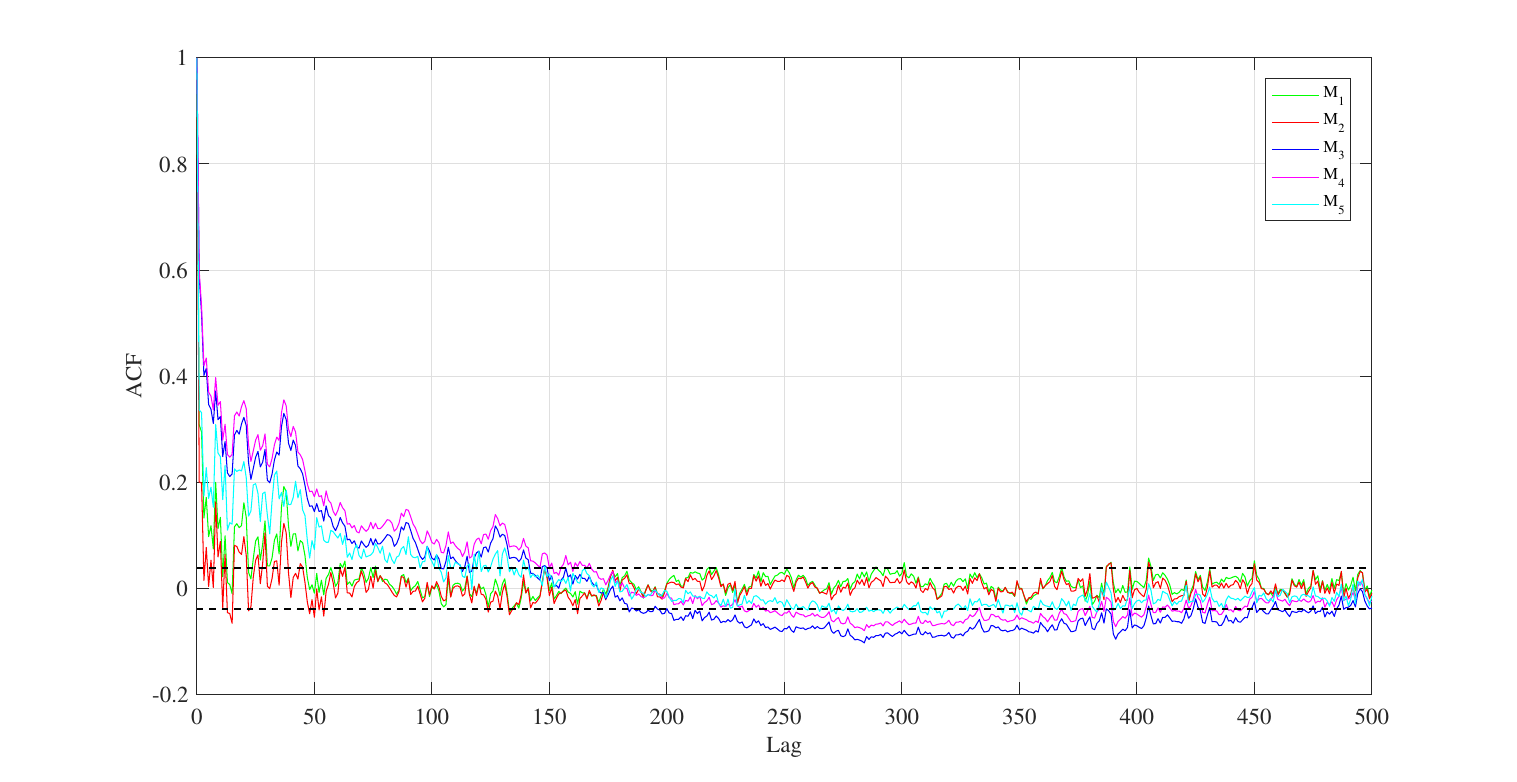}
        \caption{GEI residuals ACF under models $M_1$--$M_5$}
        \label{fig:sub4}
    \end{subfigure}
    \begin{subfigure}{0.3\textwidth}
        \centering
        \includegraphics[width=\textwidth]{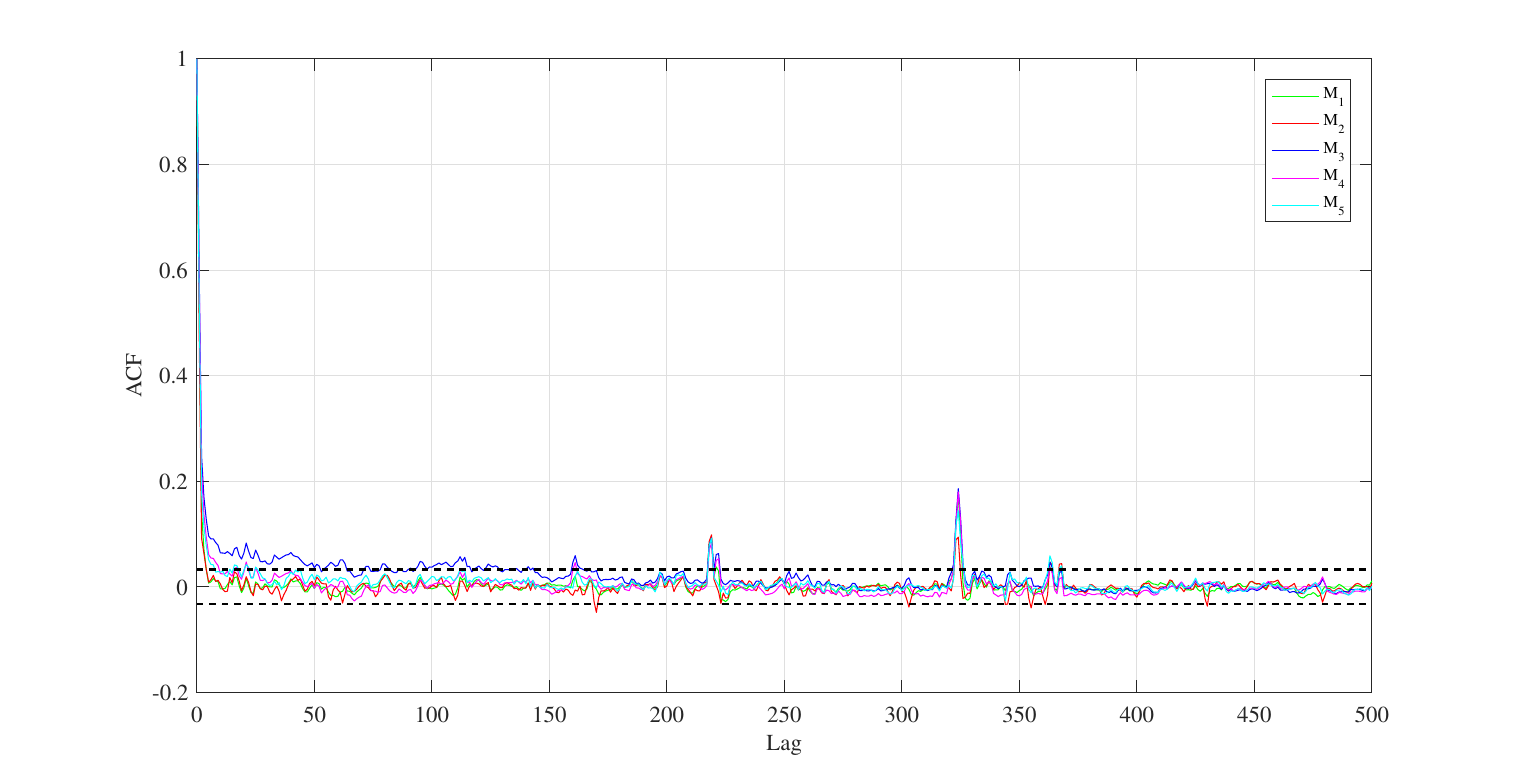}
        \caption{SSE 50 ETF residuals ACF under models $M_1$--$M_5$}
        \label{fig:sub5}
    \end{subfigure}
    \begin{subfigure}{0.3\textwidth}
        \centering
        \includegraphics[width=\textwidth]{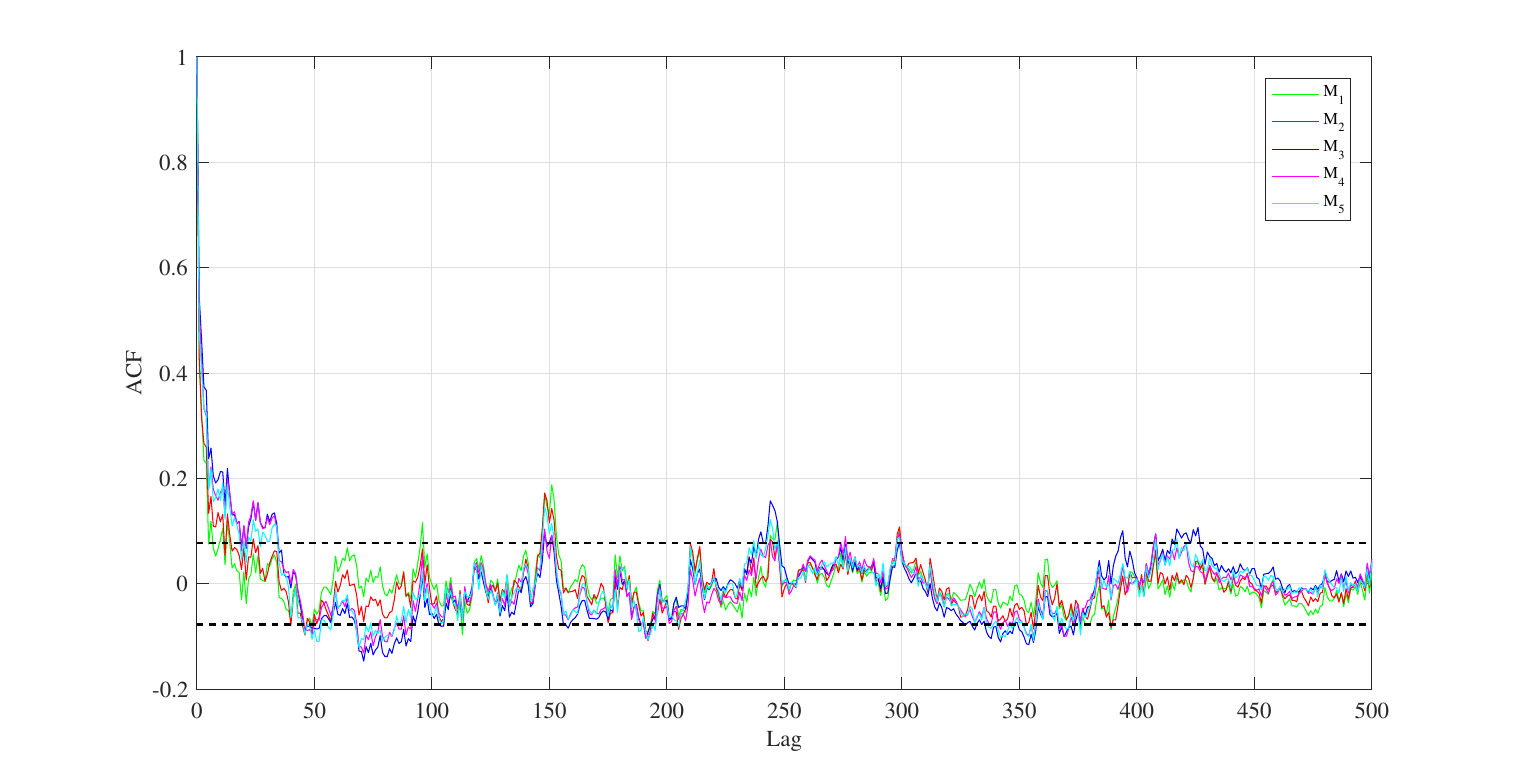}
        \caption{STAR 50 residuals ACF under models $M_1$--$M_5$}
        \label{fig:sub6}
    \end{subfigure}

    \caption{The residuals ACF of each index under model $M_1$--$M_5$}
    \label{fig:all}
\end{figure}

According to \citet{parent2022rough}, the dataset is divided into a training set, which comprises $80\%$ of the data from January 4, 2005, to March 13, 2020, and a test set, which consists of the remaining $20\%$ from March 16, 2020, to December 29, 2023.

\begin{table}[htbp]
\centering
\caption{Goodness of fit of each index under different models}
\label{tab:fitting}
\resizebox{\textwidth}{!}{%
\begin{threeparttable} 
    \begin{tabular}{lccccccccccccc}
    \toprule
    & & \multicolumn{2}{c}{SSE} & \multicolumn{2}{c}{CSI 300} & \multicolumn{2}{c}{SSE 50} & \multicolumn{2}{c}{GEI} & \multicolumn{2}{c}{SSE 50 ETF} & \multicolumn{2}{c}{STAR 50} \\
    \cmidrule(lr){3-4} \cmidrule(lr){5-6} \cmidrule(lr){7-8} \cmidrule(lr){9-10} \cmidrule(lr){11-12} \cmidrule(lr){13-14}
    & & Train& Test & Train & Test & Train & Test & Train & Test & Train & Test & Train & Test \\
    \midrule
    \multirow{2}{*}{$M_1$} & $R^2$ & $0.495$ & $0.474$ & $0.455$ & $0.383$ & $0.486$ & $0.442$ & $0.456$ & $0.299$ & $0.175$ & $0.382$ & $0.248$ & $0.059$ \\
    & RMSE & $0.00033$ & $0.00007$ & $0.00035$ & $0.00011$ & $0.00036$ & $0.00009$ & $0.00041$ & $0.00015$ & $0.00084$ & $0.00009$ & $0.00016$ & $0.00016$ \\
    \multirow{2}{*}{$M_2$} & $R^2$ & $0.521$ & $0.466$ & $0.503$ & $0.390$ & $0.530$ & $0.391$ & $0.538$ & $0.319$ & $0.206$ & $0.336$ & $0.255$ & $0.0687$ \\
    & RMSE & $0.00032$ & $0.00007$ & $0.00033$ & $0.00010$ & $0.00033$ & $0.00010$ & $0.00034$ & $0.00015$ & $0.00080$ & $0.00010$ & $0.00018$ & $0.00016$ \\
    \multirow{2}{*}{$M_3$} & $R^2$ & $0.048$ & $0.044$ & $0.040$ & $0.018$ & $0.0432$ & $0.019$ & $0.0349$ & $0.002$ & $0.000$ & $0.001$ & $0.0038$ & $0.0259$ \\
    & RMSE & $0.00042$ & $0.00009$ & $0.00105$ & $0.00141$ & $0.00046$ & $0.00012$ & $0.00051$ & $0.00018$ & $0.00096$ & $0.00125$ & $0.00018$ & $0.00019$ \\
    \multirow{2}{*}{$M_4$} & $R^2$ & $0.279$ & $0.260$ & $0.255$ & $0.205$ & $0.245$ & $0.257$ & $0.018$ & $0.001$ & $0.0362$ & $0.279$ & $0.091$ & $0.012$ \\
    & RMSE & $0.00046$ & $0.00008$ & $0.00040$ & $0.00015$ & $0.00041$ & $0.00010$ & $0.00054$ & $0.00018$ & $0.00245$ & $0.00080$ & $0.00017$ & $0.00017$ \\
    \multirow{2}{*}{$M_5$} & $R^2$ & $0.348$ & $0.200$ & $0.337$ & $0.241$ & $0.390$ & $0.196$ & $0.279$ & $0.103$ & $0.084$ & $0.239$ & $0.163$ & $0.038$ \\
    & RMSE & $0.00043$ & $0.00008$ & $0.00034$ & $0.00012$ & $0.00038$ & $0.00011$ & $0.00043$ & $0.00017$ & $0.00086$ & $0.00012$ & $0.00017$ & $0.00016$ \\
    \bottomrule    
    \end{tabular}
    \begin{tablenotes}
        \item Note: all RMSE are retained to five decimal places, and $R^2$ denotes the goodness-of-fit.
    \end{tablenotes}
\end{threeparttable} 
}
\end{table}

As shown in Figure \ref{fig:all}, it shows the results of the residual autocorrelation of each sample data in the model $M_1$--$M_5$ with a lag of 500 items. And the models $M_1$--$M_5$ are represented by green, red, blue, magenta, and cyan respectively and the two horizontal lines represent confidence intervals. Taking Figure \ref{fig:sub1} as an example, Model $M_3$ exhibits strong and persistent autocorrelation in residuals, indicating the presence of unexplained dynamic features in the data, such as nonlinear characteristics, or other factors not captured by the model. In contrast, Model $M_2$ significantly reduces residual autocorrelation, indicating that the model effectively captures the primary impact of lagged variables on the target variable.
As shown in Figures \ref{fig:sub2}-- \ref{fig:sub5}, model $M_2$  exhibits a low degree of autocorrelation, with most values falling within the confidence interval. 
This result indicates that the model $M_2$ has effectively reduced the unexplained information in the residuals, enhancing overall fitting accuracy and explanatory power. This analysis further confirms that in modeling path-dependent, volatility features play a dominant role in explaining volatility dynamics. This conclusion is consistent with the findings of \citet{parent2023investigating}. Table \ref{tab:fitting} shows the fitting performance of Models $M_1$--$M_5$ on different datasets. The results indicate that Model $M_2$ has significantly superior fitting performance on all samples (both train and test) compared with other models. By contrast, Model $M_3$ has relatively poor fitting performance. The above illustrates the contribution of the volatility feature to explaining realized volatility, and we use this feature to construct the HAR-PD-RV model at  \eqref{har-pd-rv} and use the structure of Model $M_2$ to construct the other HAR-PD models.

\begin{figure*}[!htbp]
\centering
\begin{minipage}[t]{0.45\textwidth}
\centering
\includegraphics[width=2.5in]{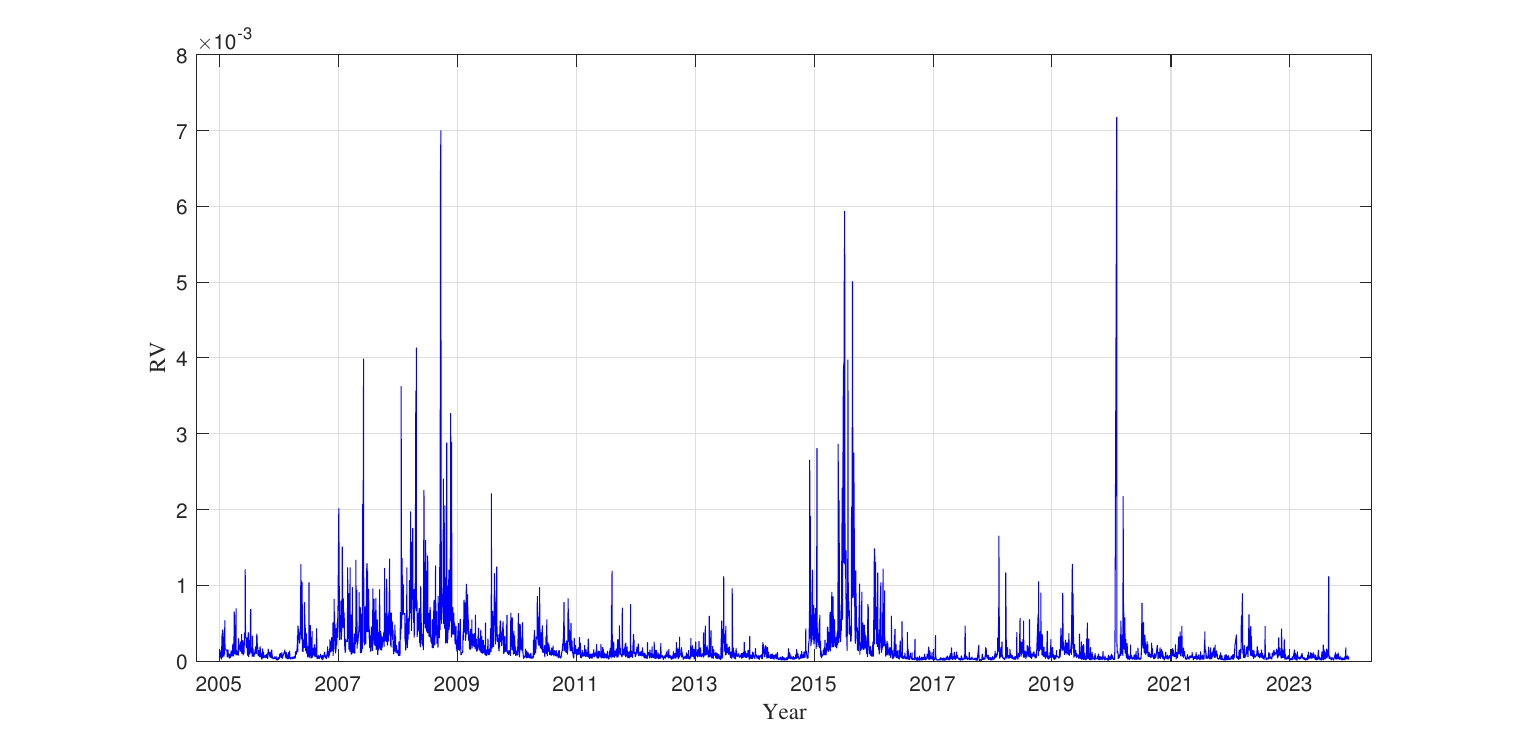}
\caption{SSE ${\rm RV}_t$}
\label{img：figure3}
\end{minipage}%
\begin{minipage}[t]{0.45\textwidth}
\centering
\includegraphics[width=2.5in]{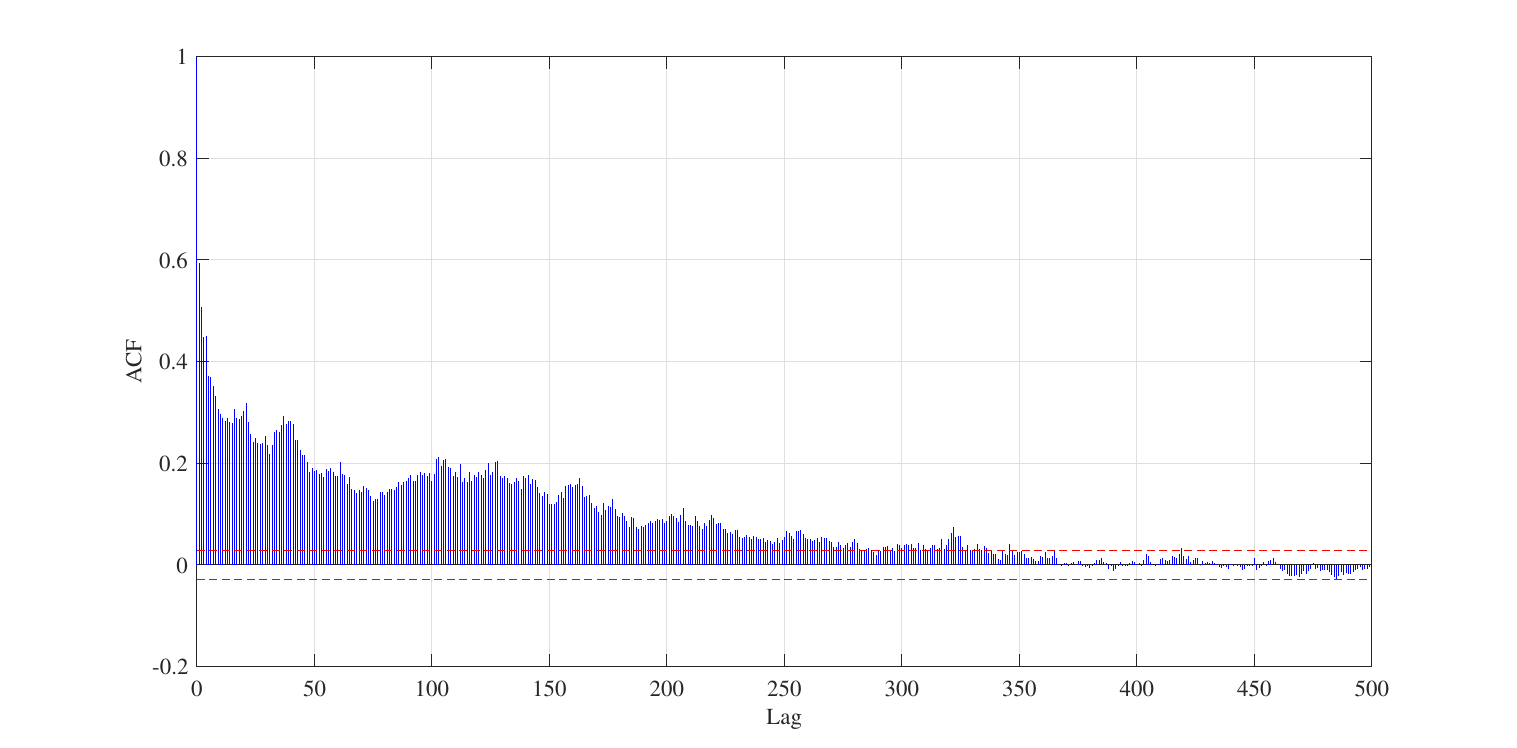}
\caption{SSE ${\rm RV}_t$ ACF}
\label{fig:RV_ACF}
\end{minipage}
\end{figure*}

\begin{table}[!htbp]
\centering
\caption{Descriptive statistics of realized volatility and other variables}
\label{tab:table1}
\resizebox{\textwidth}{!}{
\begin{threeparttable}
\begin{tabular}{lcccccccc} 
\toprule
{Variables} &
  {Mean} &
  {Standard Deviation} &
  {Skewness} &
  {Kurtosis} &
  {Maximum} &
  {Minimum} &
  {ADF} &
  {Ljung-Box(10)} \\ 
\midrule
RV & $0.201$ & $0.399$  & $7.857$   & $95.914$   & $7.175$  & $0.007$  & $-8.798^{***}$ & $7873.7^{***}$ \\
CJ & $0.849$  & $0.043$  & $16.654$   & $369.009$  & $1.131$ & $0.000$    &$-12.424^{***}$      & $137.6^{***}$ \\
CV & $0.142$  & $0.243$  & $5.828$   & $55.344$   & $3.898$  & $0.005$ &$-7.314^{***}$   & $11988^{***}$ \\      
$\mathrm{RS}^{+}$ & $0.099$  & $0.223$  & $14.884$   & $345.431$  & $6.543$ & $0.002$   &$-10.206^{***}$ & $399.7^{***}$ \\
$\mathrm{RS}^{-}$ & $0.106$  & $0.248$  & $11.156$   & $220.296$  & $7.079$ & $0.000$ &$-8.979^{***}$ & $3803.4^{***}$ \\ 
$\mathrm{REX}^{+}$ & $0.035$  & $0.076$  & $6.872$    &$76.892$  & $1.447$  & $0.000$   &$-7.619^{***}$     & $6809.6^{***}$ \\   
$\mathrm{REX}^{-}$ & $0.038$ & $0.083$ &$8.627$&$111.056$ & $1.546$ & $0.000$ &$-9.561^{***}$ &$6320.3^{***}$\\
$\mathrm{REX}^{m}$ & $0.077$  & $0.132$  & $6.379$    & $71.830$   & $2.556$  & $0.000$    & $-7.341^{***}$     & $12713^{***}$ \\ 
$\mathrm{REXQ}^{+}$ &$0.041$& $0.081$& $8.549$&$114.839$&$0.002$&$0.000$& $-8.857^{***}$&$5057.1^{***}$\\
$\mathrm{REXQ}^{-}$ &$0.039$&$0.079$&$6.930$&$72.924$&$1.45$&$0.000$&$-7.487^{***}$&$4623.4^{***}$\\
$\mathrm{REXQ}^{m}$ &$0.069$ &$0.117$ &$6.054$&$60.071$&$2.123$&$0.000$&$-7.194^{***}$&$11575^{***}$\\
\bottomrule    
\end{tabular}
\begin{tablenotes} 
   \item{Note: CV and CJ denote the continuous and jump components, respectively, derived from the underlying volatility model. The Augmented Dickey-Fuller (ADF) test examines the presence of a unit root in the series, with the null hypothesis positing that the series may be non-stationary. The Ljung-Box(10) test evaluates whether time series data exhibit autocorrelation. For ease of presentation, the mean, standard deviation, and maximum values are displayed after multiplying the original values by $10^3$. The symbols $*$, $**$, and $***$ indicate rejection of the null hypothesis at the $10\%$, $5\%$, and $1\%$ significance levels, respectively.}
\end{tablenotes}
\end{threeparttable}
}
\end{table}

\subsection{In-sample fitting results}
\begin{table}[!ht]
\centering
\scriptsize
\caption{Parameter estimation results of the HAR model family}
\label{tab:part1}

\begin{threeparttable}
\begin{tabular}{@{}lccccc@{}}
\toprule
& HAR-RV & HAR-CJ & HAR-RS & HAR-REX & HAR-REQ \\
\midrule
$\beta_0$ & $0.000^{***}$ & $0.000^{***}$ & $0.000^{***}$ & $0.000$ & $0.000$ \\
& $(0.000)$ & $(0.000)$ & $(0.000)$ & $(0.000)$ & $(0.000)$ \\
$\beta_1$ & $0.328^{***}$ & $0.437^{***}$ & $-0.111^{***}$ & $-0.494^{***}$ & $-1.059^{***}$ \\
& $(0.019)$ & $(0.023)$ & $(0.003)$ & $(0.087)$ & $(0.103)$ \\
$\beta_2$ & $0.273^{***}$ & $0.316^{***}$ & $0.814^{***}$ & $1.896^{***}$ & $2.213^{***}$ \\
& $(0.032)$ & $(0.040)$ & $(0.054)$ & $(0.204)$ & $(0.240)$ \\
$\beta_3$ & $0.266^{***}$ & $0.162^{***}$ & $0.014^{***}$ & $1.006^{***}$ & $0.442$ \\
& $(0.032)$ & $(0.043)$ & $(0.096)$ & $(0.303)$ & $(0.379)$ \\
$\beta_4$ & & $0.644^{***}$ & $0.001^{***}$ & $-0.492^{***}$ & $0.147$ \\
& & $(0.086)$ & $(0.024)$ & $(0.098)$ & $(0.105)$ \\
$\beta_5$ & & $-0.198^{***}$ & $1.298^{***}$ & $0.147^{***}$ & $1.326^{***}$ \\
& & $(0.209)$ & $(0.051)$ & $(0.216)$ & $(0.240)$ \\
$\beta_6$ & & $0.719^{***}$ & $-0.141^{***}$ & $-0.057$ & $-0.644^{*}$ \\
& & $(0.330)$ & $(0.088)$ & $(0.344)$ & $(0.362)$ \\
$\beta_7$ &&&&$-0.046$&$0.573^{***}$\\
&&&&$(0.081)$&$(0.106)$\\
$\beta_8$&&&&$1.325^{***}$&$0.884^{***}$\\
&&&&$(0.166)$&$(0.215)$\\
$\beta_9$&&&&$-0.627^{*}$&$-0.060$\\
&&&&$(0.242)$&$(0.310)$\\
$Adj. R^{2}$ & $0.406$ & $0.504$ & $0.581$ & $0.515$ & $0.517$ \\
\bottomrule
\end{tabular}
\begin{tablenotes}
\item Note: The adjusted $R^2$ value denotes the adjusted goodness-of-fit. The symbols $*$, $**$, and $***$ denote the rejection of the null hypothesis at the $10\%$, $5\%$, and $1\%$ significance levels, respectively. The values in parentheses correspond to the standard errors of the respective parameters, with all estimated results rounded to three decimal places.
\end{tablenotes}
\end{threeparttable}

\end{table}
Using the SSE data as the sample, the selected time period spans from January 4, 2005, to December 29, 2023. After data cleaning, a total of 221,568 five-minute high-frequency observations were obtained.

As illustrated in Figure \ref{img：figure3}, the time trend of realized volatility from 2005 to 2023 is presented. The vertical axis represents the magnitude of realized volatility, while the horizontal axis denotes the temporal dimension of the data. It is evident from Figure \ref{img：figure3} that during the 2008 global financial crisis, the 2016 implementation of China's first stock market circuit breaker, and the 2020 outbreak of the COVID-19 pandemic, realized volatility exhibited significant peaks, which closely align with the substantial market fluctuations induced by these major events.
Figure \ref{fig:RV_ACF} illustrates the autocorrelation function of volatility, which measures the correlation between different time points in the volatility time series. The autocorrelation structure is distinctly observable, and this characteristic is crucial for developing accurate volatility forecasting models.

\begin{table}[!htbp]
\centering
\caption{Parameter estimation results of the HAR-PD model family}
\label{tab:part2}
\begin{threeparttable}
\begin{tabular}{@{}lccccc@{}}
\toprule
& HAR-PD-RV & HAR-PD-CJ & HAR-PD-RS & HAR-PD-REX & HAR-PD-REQ \\
\midrule
$\lambda_1$ & $39.697^{***}$ & $0.002^{***}$ &$0.367^{***}$& $0.098^{***}$& $0.018^{***}$ \\
&$(0.000)$&$(0. 000)$&$(0.000)$& $(0.000)$ &$(0. 121)$\\
$\lambda_2$ &    & $24.225^{***}$               &              $15.417^{***}$&    $38.594^{***}$                  &   $9.616^{***}$                 \\
& &$(0. 000)$&$(0.000)$ &$(0.000)$&$(0.000)$\\
$\lambda_3$ & & $14.602^{***}$& $34.916^{***}$ &$47.169^{***}$ &$0.144^{***}$\\
& &$(0. 000)$&$(0. 000)$&$(0.000)$& $(0.000)$ \\
$\lambda_4$ & & & &$0.025^{***}$& $14.832^{***}$\\
& & & &$(0. 000)$&$(0.000)$\\
$\beta_0$ & $0.000^{***}$ & $0.000$ & $0.000^{*}$ & $0.000^{***}$ & $0.000$ \\
& $(0.000)$ & $(0.000)$ & $(0.000)$ & $(0.000)$ & $(0.000)$ \\
$\beta_1$ & $0.099^{***}$ & $4.848$ & $0.241^{***}$ & $0.587^{***}$ & $1.967^{***}$ \\
& $(0.008)$ & $(2.098)$ & $(0.031)$ & $(0.110)$ & $(0.564)$ \\
$\beta_2$ & $0.533^{***}$ & $-6.487^{**}$ & $-0.611^{**}$ & $-0.524^{***}$ & $-1.799^{***}$ \\
& $(0.019)$ & $(2.390)$ & $(0.044)$ & $(0.133)$ & $(0.657)$ \\
$\beta_3$ & $0.255^{***}$ & $1.637^{**}$ & $0.245^{***}$ & $0.159^{**}$ & $-0.104$ \\
& $(0.023)$ & $(0.720)$ & $(0.056)$ & $(0.047)$ & $(0.223)$ \\
$\beta_4$ & & $-0.561^{***}$ & $-0.189^{***}$ & $-0.275^{***}$ & $-13.302^{***}$ \\
& & $(0.103)$ & $(0.026)$ & $(0.077)$ & $(0.750)$ \\
$\beta_5$ & & $1.707^{***}$ & $1.127^{***}$ & $3.467^{***}$ & $12.483^{***}$ \\
& & $(0.270)$ & $(0.053)$ & $(0.167)$ & $(0.807)$ \\
$\beta_6$ &&$-0.073$&$-0.144$&$0.018$&$0.652^{*}$\\
&&$(0.466)$&$(0.096)$&$(0.272)$&$(0.331)$\\
$\beta_7$&&$0.098$&$-0.064^{**}$&$0.516^{***}$&$0.034$\\
&&$(0.032)$&$(0.020)$&$(0.081)$&$(0.103)$\\
$\beta_8$&&$1.355^{***}$&$1.593^{***}$&$2.715^{***}$ &$1.725^{***}$\\
&&$(0.053)$&$(0.051)$&$(0.179)$&$(0.237)$\\
$\beta_9$&&$-0.026$&$-0.193^{*}$ &$-0.243$&$-0.987^{**}$\\
&&$(0.053)$&$(0.098)$&$(0.300)$&$(0.367)$\\
$\beta_{10}$&&&&$-28.388^{***}$&$0.501^{***}$\\
&&&&$(2.906)$&$(0.096)$\\
$\beta_{11}$&&&&$26.528^{***}$&$2.484^{***}$\\
&&&&$(2.906)$&$(0.181)$\\
$\beta_{12}$&&&&$0.863$&$-0.690^{*}$\\
&&&&$(0.685)$&$(0.290)$\\
$Adj. R^{2}$ & $0.503$ & $0.509$ & $0.606$ & $0.523$ & $0.537$ \\
\bottomrule
\end{tabular}
\begin{tablenotes}
\item Note: The adjusted $R^2$ value denotes the adjusted goodness-of-fit. The symbols $*$, $**$, and $***$ denote the rejection of the null hypothesis at the $10\%$, $5\%$, and $1\%$ significance levels, respectively. The values in parentheses correspond to the standard errors of the respective parameters, with all estimated results rounded to three decimal places.
\end{tablenotes}
\end{threeparttable}
\end{table}

To prevent overfitting and further explain the predictive performance of the HAR-PD model family with path dependence, the LASSO method, as described in Section \ref{sec:lasso}, was employed to construct the HAR-PD model under LASSO \endnote{The specific code can be found at https://github.com/wangxiaobo018/realized-volatility}.

\begin{table}[!htbp]
\centering
\caption{Sample parameter estimation results for the LASSO-HAR-PD model family}
\label{tab:part3}

\begin{threeparttable}
\begin{tabular}{@{}lcccc@{}}
\toprule
{Model} & {Selected Variable} & {Parameter Estimate} & {Standard Deviation} & {$Adj.R^2$} \\
\midrule
\multirow{5}{*}{LASSO-HAR-PD-CJ} 
& $\overline{R}_{2,t-5}$ & $-1.477^{*}$ & $(0.759)$ & \multirow{4}{*}{$0.508$} \\
& $\overline{R}_{2,t-22}$ & $1.428^{*}$ & $(0.753)$ & \\
& $\mathrm{{PDCJ}}_{t-1}$ & $-0.533^{***}$ & $(0.102)$ & \\
& $\mathrm{\overline{PDCJ}}_{t-5}$ & $1.706^{***}$ & $(0.233)$ & \\
& $\mathrm{\overline{{PDCV}}}_{t-5}$ & $1.343^{***}$ & $(0.025)$ & \\
\midrule
\multirow{4}{*}{LASSO\text{-}HAR\text{-}PD\text{-}RS}
& $\overline{R}_{2,t-5}$ & $-3.819^{***}$ & $(0.322)$ & \multirow{4}{*}{$0.602$} \\
& $\overline{R}_{2,t-22}$ & $3.576^{***}$ & $(0.297)$\\
& $\mathrm{{PDRS}}_{t-1}^{+}$ & $-0.136^{***}$ & $(0.025)$\\
& $\mathrm{\overline{{PDRS}}}_{t-5}^{+}$ & $0.973^{***}$ & $(0.046)$ & \\
& $\mathrm{\overline{{PDRS}}}_{t-5}^{-}$ & $1.073^{***}$ & $(0.038)$ & \\
& $\mathrm{\overline{{PDRS}}}_{t-22}^{-}$ & $0.264^{***}$ & $(0.064)$&\\
\midrule
\multirow{7}{*}{LASSO\text{-}HAR\text{-}PD\text{-}REX} 
& $R_{2,t-1}$ & $0.791^{***}$ & $(0.156)$ & \multirow{7}{*}{$0.522$} \\
& $\overline{R}_{2,t-5}$ & $0.282^{***}$ & $(0.020)$ & \\
& $\mathrm{{PDREX}}_{t-1}^{+}$ & $-0.234^{***}$ & $(0.075)$ & \\
& $\mathrm{\overline{{PDREX}}}_{t-5}^{+}$ & $33.688^{**}$ & $(3.040)$ & \\
& $\mathrm{{PDREX}}_{t-1}^{-}$ & $0.548^{***}$ & $(0.079)$ & \\
& $\mathrm{\overline{{PDREX}}}_{t-5}^{-}$ & $2.529^{***}$ & $(0.148)$ & \\
& $\mathrm{{{PDREX}}}_{t-1}^{m}$ & $-34.436^{***}$ & $(3.104)$ &\\
& $\mathrm{\overline{{PDREX}}}_{t-5}^{m}$ & $33.689^{***}$ & $(3.041)$ & \\
\midrule
\multirow{7}{*}{LASSO\text{-}HAR\text{-}PD\text{-}REQ} 
& $R_{2,t-1}$ & $5.456^{***}$ & $(1.389)$ & \multirow{7}{*}{$0.533$} \\
& $\overline{R}_{2,t-5}$ & $-5.437^{***}$ & $(1.382)$ & \\
& $\mathrm{{PDREQ}}_{t-1}^{+}$ & $-1.388^{***}$ & $(0.228)$ & \\
& $\mathrm{\overline{{PDREQ}}}_{t-5}^{+}$ & $1.271^{***}$ & $(0.748)$ & \\
& $\mathrm{\overline{{PDREQ}}}_{t-5}^{-}$ & $-1.900^{***}$ & $(0.194)$ & \\
& $\mathrm{\overline{{PDREQ}}}_{t-22}^{-}$ & $-1.388^{**}$ & $(0.228)$ & \\
& $\mathrm{{{PDREQ}}}_{t-1}^{m}$ & $5.303^{***}$ & $(0.073)$ &\\
& $\mathrm{\overline{{PDREQ}}}_{t-5}^{m}$ & $2.301^{***}$ & $(0.152)$ & \\
\bottomrule
\end{tabular}
\begin{tablenotes}
\item Note: The adjusted $R^2$ value denotes the adjusted coefficient of determination. The symbols $*$, $**$, and $***$ denote the rejection of the null hypothesis at the $10\%$, $5\%$, and $1\%$ significance levels, respectively. The values in parentheses correspond to the standard errors of the respective parameters, with all estimated results rounded to three decimal places.
\end{tablenotes}
\end{threeparttable}
\end{table}
Table \ref{tab:table1} presents the statistical description of volatility based on five-minute high-frequency data, with values of all variables except Ljung-Box(10) test rounded to three decimal places. The distribution of volatility exhibits significant right skewness, indicating a higher frequency of large volatility values. Additionally, the kurtosis value is 95.914, suggesting a highly peaked distribution with fat tails, which implies the potential presence of extreme values.
The ADF test statistics for all variables are negative, and the null hypothesis of a unit root is rejected at the $1\%$ significance level, indicating that all time series variables are stationary.
Taking the SSE as an example, Tables \ref{tab:part1} and \ref{tab:part2} present the parameter estimation results of the HAR and HAR-PD model families. In terms of adjusted $R^2$, the HAR-RS model demonstrates the highest adjusted goodness-of-fit among the HAR family models, which is consistent with the volatility prediction model findings reported by \citet{zhang2023volatility}.
As shown in Table \ref{tab:part2}, the HAR-PD-RV model exhibits the adjusted $R^2$ of 0.503, with parameters $\beta_0$, $\beta_1$, and $\beta_2$ being statistically significant at the $1\%$ level, substantially outperforming the benchmark HAR-RV model's adjusted $R^2$ of 0.406. A notable improvement in the adjusted $R^2$ is also observed when comparing the HAR-CJ model with the HAR-PD-CJ model. The HAR-PD model family demonstrates higher significance levels compared to the benchmark HAR model family. The HAR-PD-RS model achieves the highest adjusted $R^2$ with multiple significant parameters. Similarly, the empirical quantile decomposition-based HAR-PD-REQ model shows superior the adjusted $R^2$ compared to the benchmark HAR-REX model. Overall, the HAR-PD model family consistently demonstrates higher adjusted $R^2$ values than the HAR model family.
Table \ref{tab:part3} presents the model variables selected through LASSO, where variables $R_{2,t-1}$ and $\overline{R}_{2,t-5}$ were consistently selected across all models in the LASSO-HAR-PD family, further substantiating the contribution of volatility features to the models.
The information criterion of the HAR model family and the LASSO-HAR-PD model family shown in the Table \ref{harAIC} shows that a balance has been achieved between the complexity and fit of the LASSO-HAR-PD model family. The improved model does not achieve a better fit by increasing complexity. And the HAR-REQ model constructed by empirical quantiles has smaller AIC and BIC than the HAR-REX model.
\begin{table}[htbp]
\centering
\label{tab:combined_tables}
\begin{threeparttable}
\centering
\caption{Information criteria for the HAR and HAR-PD model family}\label{harAIC}
\begin{tabular}{@{}lcc@{}}
\toprule
Model & {AIC} & {BIC} \\
\midrule
HAR-RV & $-64062.49$ & $-64030.30$ \\
HAR-CJ & $-61909.70$ & $-61858.94$ \\
HAR-RS & $-63022.55$ & $-62971.09$ \\
HAR-REX & $-62253.74$ & $-62182.94$ \\
HAR-REQ & $-62468.68$ & $-62193.92$ \\
HAR-PD-RV & $-62403.08$ & $-62370.92$ \\
LASSO-HAR-PD-CJ & $-62986.16$ & $-62947.54$ \\
LASSO-HAR-PD-RS & $-63449.58$ & $-63410.97$ \\
LASSO-HAR-PD-REX & $-63008.93$ & $-62951.00$ \\
LASSO-HAR-PD-REQ & $-62823.72$ & $-62759.39$ \\
\bottomrule
\end{tabular}
\begin{tablenotes}
\item Note: AIC and BIC denote the Akaike information criterion and the Bayesian information criterion, respectively.
\end{tablenotes}
\end{threeparttable}
\end{table}

\subsection{Out-of-sample prediction results}
\subsubsection{Rolling time window prediction and Out-of-sample prediction evaluation methods}\label{sec:6}

The primary objective of volatility model research is to enhance out-of-sample prediction capabilities. In this study, we employed the Rolling Window Forecasting method to generate predictions for the HAR-PD model family.
The specific procedure is outlined as follows:
\begin{enumerate}[(i)]
   \item The sample data are divided into in-sample (trian sample) and out-of-sample (test sample). Following prior studies \citep{haugom2014forecasting,feng2024out}, out-of-sample the length of the out-of-sample is generally 300, 600, 1000.  In this paper, we will use the out-of-sample of length 1000 in the robustness test. 
   After data cleaning, the in-sample is a fixed window of 4016 days from February 2, 2005, to July 6, 2021, and the out-of-sample is the last 600 trading days.
   \item Data from $t=1, 2, \cdots, N=4295$ was selected as the total  sample. In this case, when predicting one-step-ahead into the future, with a fixed window of 4,016 observations ($H=4016$), the predicted sample is the last 600 days ($M=600$), similarly when predicting five-step-ahead into the future, the predicted sample is 596 days ($M=596$), and when predicting twenty-two-step-ahead into the future, the predicted sample is 579 days ($M=579$).
   \item The estimated sample intervals are shifted backward with a constant length of the fixed window, and the model's parameters are re-estimated for each backward projection of future volatility to ensure that the sample size used for estimation remains constant.
\end{enumerate}
To comprehensively validate the advantages of the path-dependent structure beyond merely comparing the HAR model family, several researchers \citep{moreno2024deepvol,hansen2005forecast} have incorporated additional prediction models as benchmark comparisons. Thus, in this context, we utilize the GARCH model \citep{BOLLERSLEV1986307} and its extensions as benchmark models.

\begin{table}[ht]
\centering
\caption{Out-of-sample forecast loss value of SSE for the one-step-ahead predictions}
\label{tab:table4}
\resizebox{\textwidth}{!}{
\begin{threeparttable}

\begin{tabular}{lcccccc}
\toprule
{Model} & {MSE} & {MAE} & {HMSE} & {HMAE} & {QLIKE} \\ \midrule
GARCH & $6.442 \times 10^{-9}$ & $4.657 \times 10^{-5}$ & $1.975$ & $0.937$ & $-8.608$ \\
EGARCH & $5.692 \times 10^{-9}$ & $4.657 \times 10^{-5}$ & $1.975$ & $0.937$ & $-8.629$ \\
TGARCH & $5.259 \times 10^{-9}$ & $4.841 \times 10^{-5}$ & $1.942$ & $0.995$ & $-8.657$ \\
GJRGARCH & $8.954 \times 10^{-9}$ & $6.429 \times 10^{-5}$ & $4.726$ & $1.553$ & $-8.602$ \\
HAR-RV & $7.025 \times 10^{-9}$ & $4.887 \times 10^{-5}$ & $1.924$ & $1.001$ & $-8.609$ \\
HAR-CJ & $6.180 \times 10^{-9}$ & $4.133 \times 10^{-5}$ & $1.006$ & $0.743$ & $-8.648$ \\
HAR-RS & $4.427 \times 10^{-9}$ & $3.339 \times 10^{-5}$ & $0.836$ & $0.578$ & $-8.724$ \\
HAR-REX & $5.081 \times 10^{-9}$ & $3.432 \times 10^{-5}$ & $0.587$ & $0.552$ & $-8.697$ \\
HAR-REQ & $5.056 \times 10^{-9}$ & $3.344 \times 10^{-5}$ & $0.526$ & $0.524$ & $-8.698$ \\
HAR-PD-RV& $6.858 \times 10^{-9}$ & $3.955 \times 10^{-5}$ & $0.802$ & $0.574$ & $-8.607$ \\
HAR-PD-CJ& $4.693 \times 10^{-9}$ & $3.324 \times 10^{-5}$ & $0.780$ & $0.536$ & $-8.726$ \\
HAR-PD-RS & \textbf{$\underline{\mathbf{4.185 \times 10^{-9}}}$} & $3.446 \times 10^{-5}$ & $1.035$ & $0.625$ & $-8.652$ \\
HAR-PD-REX& $4.923 \times 10^{-9}$ & $3.286 \times 10^{-5}$ & $0.522$ & $0.511$ & $-8.700$ \\
HAR-PD-REQ & $4.855 \times 10^{-9}$ & \textbf{$\underline{\mathbf{3.116 \times 10^{-5}}}$} & \textbf{$\underline{\mathbf{0.450}}$} & \textbf{$\underline{\mathbf{0.467}}$} & $-8.707$ \\
LASSO-HAR-PD-CJ& $5.038 \times 10^{-9}$ & $3.277 \times 10^{-5}$ & $0.462$ & $0.483$ & $-8.701$ \\
LASSO-HAR-PD-RS& $4.391 \times 10^{-9}$ & $3.152 \times 10^{-5}$ & $0.671$ & $0.491$ & \textbf{$\underline{\mathbf{-8.733}}$} \\
LASSO-HAR-PD-REX & $5.022 \times 10^{-9}$ & $3.463 \times 10^{-5}$ & $0.625$ & $0.571$ & $-8.692$ \\
LASSO-HAR-PD-REQ & $4.981 \times 10^{-9}$ & $3.317 \times 10^{-5}$ & $0.535$ & $0.520$ & $-8.698$ \\ \bottomrule
\end{tabular}

\begin{tablenotes}
\item Note: The bold values in the table indicate the minimum under the corresponding loss function, while the underlined values denote the optimal models identified by the MCS test at a significance level of 0.25.
\end{tablenotes}

\end{threeparttable}
}
\end{table}

For the evaluation of forecasting accuracy, some scholars, such as \citet{hyndman2006another}, \citet{armstrong2010long}, and \citet{hansen2005forecast}, have suggested employing multiple loss functions. In this study, five loss functions are utilized: 
\begin{align*}
& {L_1:}\, \mathrm{MSE} = M^{-1}\sum_{m=H+1}^{H+M} (RV_m-\hat{\sigma}_m^2)^2 \,,~~
{L_2:}\,\text{MAE} = M^{-1}\sum_{m=H+1}^{H+M}\mid RV_m-\hat{\sigma}_m^2\mid \,, \\
&{L_3:}\, \mathrm{HMSE} = M^{-1}\sum_{m = H+1}^{H+M} \left( 1-\frac{\hat{\sigma}_m^2}{RV_m} \right)^2 \,, ~~ {L_4:}\,\text{HMAE} = M^{-1}\sum_{m=H+1}^{H+M}\left|1-\frac{\hat{\sigma}_m^2}{RV_m}\right| \,, \\
& {L_5:}\,\mathrm{QLIKE} = M^{-1}\sum_{m=H+1}^{H+M}\left\{\ln(\hat{\sigma}_m^2)+\frac{RV_m}{\hat{\sigma}_m^2}\right\} \,.
\end{align*}
Here, $\hat{\sigma}_m^2$ represents the out-of-sample predicted value, while $RV_m$ denotes the realized market volatility at time $m$. Additionally, $M$ corresponds to the forecasting sample size.

When utilizing the aforementioned loss functions to evaluate model accuracy, relying solely on a single loss function for comparison is insufficient. If the prediction loss of one model is lower than that of another, it is directly concluded that the first model has higher predictive accuracy. However, this judgment is limited and cannot be easily generalized to other datasets or scenarios involving different loss functions. In particular, extreme outliers in the data may significantly impact the calculation of loss functions, potentially leading to misleading evaluations of the volatility model’s performance. 
Building on this foundation, some scholars \citep{white2000reality, diebold2002comparing, hansen2005forecast} have proposed different methods for evaluating the out-of-sample predictive ability of models. However, the most widely applied approach is the MCS method \endnote{The specific MCS inspection process is shown in the \ref{sec:MCS}.}, introduced by \citet{hansen2011model}. The MCS test is employed to select the optimal forecasting model at a given confidence level. The MCS test statistics, $T_R$ and $T_{max}$, along with the corresponding $p$-values, are obtained through 5000 bootstrap simulations. In subsequent chapters, we will use the MCS test to test the accuracy of out-of-sample forecast results.

\subsubsection{Out-of-sample volatility forecasting}\label{sec:7}
Table \ref{tab:table4} shows that with SSE as the dataset and a rolling time window, we predict the last 600 days of the dataset and present the one-step-ahead prediction loss values under various loss functions (MSE, MAE, HMSE, HMAE, and QLIKE). The HAR model family has much lower prediction loss values than the GARCH model family, which is consistent with \citet{luo2020realized}. The HAR-PD model family performs well in all five loss functions and maintains low loss values even under the LASSO constraint. Notably, in Tables \ref{tab:table5} and \ref{tab:table6}, when predicting the future five and twenty-two-step-ahead, the HAR-PD-RS and LASSO-HAR-PD-RS models have significantly lower loss values than the other models. In addition, the HAR-REQ model, which defines the threshold based on the empirical quantile, outperforms the HAR-REX model, and the HAR-PD-REQ model has the lowest loss values under the MAE, HMSE, and HMAE loss functions.

In order to further test the prediction accuracy of the model, according to the suggestion of \cite{amendola2020model,hansen2003choosing}, we use the MCS test based on the QLIKE loss function and under the MSE loss function to evaluate the out-of-sample prediction accuracy. Based on the MCS test results \endnote{ Owing to space limitations, the specific results see \ref{sec:MCStest}.}, the following conclusions can be drawn:
\begin{enumerate}[(i)]
    \item As shown in Tables \ref{tab:table7}, \ref{tab:table8}, and \ref{tab:table9} for the MCS test results under the QLIKE and MSE loss functions, the HAR model family has more models passing the test than the GARCH model family. This is consistent with the findings of previous studies \cite{vortelinos2017forecasting,qu2018modeling,gong2019modeling}. Moreover, the HAR-PD model family, with its improved path-dependent structure, outperforms the benchmark HAR model family across various loss functions, achieving a notable enhancement in predictive accuracy.
    \item Our HAR-REQ model passes the MCS test more frequently than the benchmark HAR-REX. Similarly, HAR-PD-REQ and LASSO-HAR-PD-REQ also pass the MCS test more frequently than HAR-REQ. As shown in Tables \ref{tab:table7}, \ref{tab:table8}, and \ref{tab:table9}, the LASSO-HAR-PD-RS model demonstrates the highest predictive accuracy multiple times under the MCS test.
    
\end{enumerate}
In summary, the HAR-PD model family demonstrates superior out-of-sample predictive accuracy compared to the HAR model family, further confirming the advantages of volatility forecasting from a path-dependent perspective.

\subsection{Robustness testing}
In volatility forecasting models, out-of-sample forecasting accuracy is more important than in-sample forecasting accuracy. In order to further validate the forecasting accuracy of the new model, we will conduct a robustness test for out-of-sample forecasting.
First, in Section \ref{sec:7}, we adopted various loss functions to evaluate the prediction accuracy of the model based on the MCS test. To further test the robustness of the model, the out-of-sample \(R^2\) test proposed by \citet{campbell2008predicting} will be implemented.

In Section \ref{sec:6}, a rolling window prediction was conducted using the last 600 observations as the prediction sample, following the methodology outlined in \cite{yang2015realized}. For robustness analysis, a shorter rolling window was selected compared to the out-of-sample forecast, which allowed for a longer prediction interval. To evaluate the model's performance on actual unknown data, this study employed the first 3,616 observations as the rolling window length and the subsequent 1,000 observations as the prediction window. After data cleaning, the in-sample spanned from January 18, 2005, to November 4, 2019, while the out-of-sample covered the period from November 5, 2019, to December 24, 2023.

Secondly in order to assess the model's forecasting ability on other data, new data will be used here for forecasting. Section \ref{sec:6} uses the SSE, while the robustness test uses 970 days of CSI 300 data from January 2, 2020 to December 29, 2023. The data is divided into a 670-day in-sample from January 2, 2020 to October 11, 2022 and a 300-day out-of-sample from October 12, 2022 to December 29, 2023.

Finally, to comprehensively evaluate the superiority of the extended model, additional benchmark models were introduced for comparative analysis beyond the benchmark HAR model family, following the approaches of \cite{qu2018modeling} and \cite{moreno2024deepvol}, who incorporated various traditional benchmark models to demonstrate the performance advantages of their extended models.

\subsubsection{Out-of-sample $R^2$ test}
The out-of-sample $R^2$ (commonly denoted as $R^2_{\mathrm{oos}}$(\%)) is used to evaluate the model by comparing the predictive error of the forecasting model with that of the benchmark model. If the model provides better forecasts than the benchmark model, its $R^2_{\mathrm{oos}}$(\%) will be positive. The formula is given as follows:
$$
R^2_{\mathrm{oos}}(\%)=1-\dfrac{\sum\limits_{t=1}^{M}(\mathrm{RV}_t-\mathrm{RV}_t^j)^2}{\sum\limits_{t=1}^M(\mathrm{RV}_t-\mathrm{RV}_t^0)^2}, \quad j=\mathrm{Model}(1, 2, 3, \cdots, 18)\,.
$$
Here, $\mathrm{RV}_t$ represents the actual realized volatility, $\mathrm{RV}^j_t$ denotes the out-of-sample predicted value from the forecasting model, and $\mathrm{RV}^0_t$ refers to the out-of-sample predicted value from the benchmark model.
If $R^2_{\mathrm{oos}}(\%)$ > 0, it indicates that the model's mean squared predictive error (MSPE-adjusted) is lower than that of the benchmark model, demonstrating positive predictive capability. If $R^2_{\mathrm{oos}}(\%)$ = 0, the forecasting performance of the model is identical to that of the benchmark model; if $R^2_{\mathrm{oos}}(\%)< 0$ , the model performs worse than the benchmark model. The method for calculating the adjusted MSPE is detailed in \citet{clark2007approximately}.

\subsubsection{Different rolling window lengths}

\begin{table}[htb]
\centering
\caption{$R^2_{\mathrm{oos}}$(\%) test results based on a 1,000-day predicting sample}
\label{tab:table11}
\begin{threeparttable}
\begin{tabular}{lccc}
\toprule
Model & $R^2_{\mathrm{oos}}(\%)$ &{MSPE}-adjust & $p$-value\\\midrule
GARCH & $-0.172$ & $3.881 \times 10^{-9}$ & $0.265$ \\ 
EGARCH & $-0.633 $ & $2.571 \times 10^{-9}$ & $0.852$ \\ 
TGARCH& $-0.164$ & $4.219 \times 10^{-9}$ & $0.091$ \\ 
GJRGARCH & $-0.136$ & $4.779 \times 10^{-9}$ & $0.226$ \\ 
HAR-CJ & $0.187$ & $2.154 \times 10^{-8}$ & $0.084$ \\ 
HAR-RS & $0.249$ & $3.104 \times 10^{-8}$ & $0.100$ \\ 
HAR-REX & $0.187$ & $2.153 \times 10^{-8}$ & $0.085$ \\ 
HAR-REQ & $0.187$ & $2.153 \times 10^{-8}$ & $0.085$ \\ 
HAR-PD-RV & $0.255$ & $2.738 \times 10^{-8}$ & $0.079$ \\ 
HAR-PD-CJ & $0.262$ & $2.761 \times 10^{-8}$ & $0.071$ \\ 
HAR-PD-RS & $0.256$ & $3.437 \times 10^{-8}$ & $0.105$ \\ 
HAR-PD-REX & {${0.263}$} & $2.636 \times 10^{-8}$ & $0.062$ \\ 
HAR-PD-REQ & $0.259$ & $2.627 \times 10^{-8}$ & $0.063$ \\ 
LASSO-HAR-PD-CJ & $0.183$ & $2.140 \times 10^{-8}$ & $0.085$ \\ 
LASSO-HAR-PD-RS & \textbf{$\mathbf{0.291}$} & $3.331 \times 10^{-8}$ & $0.102$ \\ 
LASSO-HAR-PD-REX & $0.264$ & $2.632 \times 10^{-8}$ & $0.065$ \\ 
LASSO-HAR-PD-REQ& $0.256$ & $2.732 \times 10^{-8}$ & $0.064$ \\
\bottomrule
\end{tabular}
\begin{tablenotes}
\item {Note: Bold values indicate the highest  $R^2_{\mathrm{oos}}(\%)$ among the models.}
\end{tablenotes}
\end{threeparttable}
\end{table}
As shown in Table \ref{tab:table11}, all models in the HAR-PD family exhibit an $R^2_{\mathrm{oos}}(\%)$ value significantly greater than zero, indicating their strong effectiveness in out-of-sample forecasting. Specifically, when comparing the HAR and HAR-PD model families, the LASSO-HAR-PD-RS model demonstrates the highest $R^2_{\mathrm{oos}}(\%)$ value in out-of-sample forecasting. 

Furthermore, in Table \ref{tab:table11}, the $R^2_{\mathrm{oos}}(\%)$ values of the LASSO-HAR-PD model family and HAR-PD model family are also higher than those of the HAR model family, highlighting the overall improvement in predictive accuracy of the HAR-PD model family over traditional HAR model family. These findings further confirm that volatility forecasting models constructed from a path-dependence perspective exhibit superior predictive accuracy compared to benchmark models.

\subsubsection{Different sample data}

\begin{table}[htb]
\centering
\caption{Out-of-sample forecast loss value of CSI 300 for the one-step-ahead predictions}
\label{tab:table12}
\resizebox{\textwidth}{!}{
\begin{threeparttable}

\begin{tabular}{lccccc}
\toprule
Model & MSE & MAE & HMSE & HMAE & QLIKE \\ 
\midrule
GARCH & $1.120 \times 10^{-8}$ & $5.216 \times 10^{-5}$ & $1.743$ & $0.786$ & $-8.424$ \\
EGARCH & $2.818 \times 10^{-8}$ & $7.593 \times 10^{-5}$ & $8.771$ & $1.378$ & $-8.418$ \\
TGARCH & $1.436 \times 10^{-8}$ & $6.703 \times 10^{-5}$ & $3.762$ & $1.209$ & $-8.438$ \\
GJRGARCH & $1.104 \times 10^{-8}$ & $5.122 \times 10^{-5}$ & $1.585$ & $0.752$ & $-8.421$ \\

HAR-RV & $1.228 \times 10^{-8}$ & $5.466 \times 10^{-5}$ & $3.409$ & $0.865$ & $-8.364$ \\ 
HAR-CJ & $8.565 \times 10^{-9}$ & $3.831 \times 10^{-5}$ & $0.477$ & ${0.496}$ & $-8.536$ \\ 
HAR-RS & $7.919 \times 10^{-9}$ & $3.905 \times 10^{-5}$ & $1.003 $ & $0.571$ & $-8.553$ \\ 
HAR-REX & $9.162 \times 10^{-9}$ & $4.421 \times 10^{-5}$ & $0.800 $ & $0.666$ & $-8.513$ \\ 
HAR-REQ & $8.866 \times 10^{-9}$ & \textbf{$\mathbf{3.818 \times 10^{-5}}$} & \textbf{$\mathbf{0.433}$} & \textbf{$\mathbf{0.484}$} & $-8.525$ \\ 
HAR-PD-RV & $9.513 \times 10^{-9}$ & $4.577 \times 10^{-5}$ & $1.241 $ & $0.731$ & $-8.501$ \\ 
hHAR-PD-CJ & $8.422 \times 10^{-9}$ & $4.226 \times 10^{-5}$ & $0.688$ & $0.604$ & $-8.529$ \\ 
HAR-PD-RS &$7.774 \times 10^{-9}$ & $3.886\times 10^{-5}$ & $1.771$ & $0.574$ & \textbf{$\mathbf{-8.555}$}\\
HAR-PD-REX & $8.064 \times 10^{-9}$ & $4.062 \times 10^{-5}$ & $0.640 $ & $0.593$ & $-8.535$ \\ 
HAR-PD-REQ& \textbf{$\mathbf{7.740 \times 10^{-9}}$} & $3.956 \times 10^{-5}$ & $0.602 $ & $0.569$ & $-8.554$ \\ 
LASSO-HAR-PD-CJ & $8.427 \times 10^{-9}$ & $4.250 \times 10^{-5}$ & $0.703 $ & $0.612$ & $-8.528$ \\ 
LASSO-HAR-PD-RS & $7.638 \times 10^{-9}$ & $3.757 \times 10^{-5}$ & $0.956$ & $0.590$ & $-8.597$ \\
LASSO-HAR-PD-REX & $7.741 \times 10^{-9}$ & $3.951 \times 10^{-5}$ & $0.599 $ & $0.567$ & $-8.553$ \\ 
LASSSO-HAR-PD-REQ & $8.283 \times 10^{-9}$ & $4.140 \times 10^{-5}$ & $0.647 $ & $0.600$ & $-8.526$ \\

\bottomrule
\end{tabular}
\begin{tablenotes}
\item Note: The bold values in the table indicate the minimum under the corresponding loss function, while the underlined values denote the optimal models identified by the MCS test at a significance level of 0.25.
\end{tablenotes}
\end{threeparttable}
}
\end{table}
Table \ref{tab:table12} shows that for one-step-ahead predictions, the HAR-REQ and HAR-PD-REQ models have lower loss values across two different loss functions. The HAR-PD-RS model achieves the lowest loss value under the QLIKE loss function. As shown in Table \ref{tab:table13} for the five-step-ahead forecast and Table \ref{tab:table14} for the twenty-two-step-ahead prediction, the proposed HAR-REQ model and its improved HAR-PD-REQ variant outperform other models in volatility forecasting under the MAE, HMSE, and HMAE loss functions. Table \ref{tab:table15} presents the MCS test based on the above loss functions. The test results indicate that the HAR-PD model family and the LASSO-HAR-PD model family have passed the test many times.

\section{Conclusion}\label{sec:5}

This study systematically investigated volatility modeling and forecasting based on high-frequency data. First, an innovative approach to modeling and forecasting volatility from a path-dependence perspective was proposed. Specifically, the fundamental theory of path-dependent was introduced, and in combination with other foundational volatility decomposition theories, a more flexible and accurate volatility model was constructed.
Furthermore, a new volatility model was developed from the path dependence perspective to enhance the in-sample fitting and predictive capabilities of mainstream HAR model families. In a high-frequency data environment, the incorporation of path-dependent features allows the new model to better capture the dynamic behavior of stock market volatility. The construction process of the new model is described in detail, including its mathematical formulation, parameter estimation methods, and theoretical analysis.

To validate the effectiveness of the new model, an empirical analysis was conducted using five-minute high-frequency trading data from the SSE. Various evaluation methods and testing techniques were employed to comprehensively assess the new model, including in-sample fitting, out-of-sample forecasting, rolling window evaluation, and multiple statistical tests. The results indicated that the HAR-PD model family demonstrated superior predictive performance across all evaluations.

In out-of-sample forecasting, the HAR-PD model family, constructed based on path dependence, consistently outperformed the original HAR model family, significantly improving prediction accuracy. This result remained significant across different datasets, indicating the strong robustness of the new model. Notably, under the rolling window and out-of-sample evaluation methods, the new model exhibited sustained superiority. It is worth mentioning that among all models examined in this study, the HAR-PD-RS model achieved the highest prediction accuracy in forecasting daily volatility, while the HAR-PD-REQ model demonstrated superior predictive capability in medium- and long-term (five-step-ahead prediction and twenty-two-step-ahead prediction) volatility forecasting.

In conclusion, the HAR-PD model family constructed in this study offers new possibilities for capturing the complexity of stock market volatility. The volatility model, developed by integrating the HAR model family with a path-dependence perspective, significantly enhances the accuracy and reliability of volatility forecasting. The findings of this study not only enrich the theoretical literature on volatility modeling but also provide valuable tools for practical applications, aiding investors and risk managers in better understanding and forecasting market fluctuations. Since the HAR model family and the HAR-PD model family lack sufficient ability to capture the nonlinear fluctuation characteristics in high-frequency data, our subsequent research will focus on improving the nonlinear prediction performance of such models.

\section*{Declaration}
The authors declare no conflicts of interest.

\theendnotes

\bibliographystyle{elsarticle-num-names} 
\bibliography{reference}
\newpage
\appendix

\section{The procedure of MCS test}\label{sec:MCS}
The MCS testing procedure is as follows:
\begin{enumerate}[(i)]
    \item Given the existence of $m_0 = 18$ distinct volatility forecasting models, these models collectively form the set $M_0$, defined as $M_0 = {1, 2, \cdots, m_0}$. Each model provides daily market volatility estimates for the subsequent $M$ days, denoted as $\hat{\sigma}^2_m$ ($m = H+1, \cdots, H+M$). Based on the five loss functions defined in this study ($L_1, L_2, \cdots, L_5$), the loss function value for each prediction is computed as $L_{i, j, m}$, where $i=1, 2, \cdots, 5$; $j=1, 2, \cdots, m_0$; and $m=H+1, \cdots, H+M$. For any two models $u$ and $v$ ($u, v \in M_0$), their relative loss function value is calculated as $d_{i, uv, m} = L_{i, u, m} - L_{i, v, m}$.

   \item  A superior model set $M^*$ is defined as A superior model set \( M^* \) is defined as:
\[
M^* \equiv \{ u \in M_0 : \mathbb{E}(d_{i, uv, m}) \leq 0, \text{ for all } v \in M_0 \}
\] The MCS testing process involves conducting a series of significance tests within the set $M_0$ to eliminate models with inferior predictive ability. In each test iteration, the null hypothesis assumes that the two models possess identical predictive ability, expressed as:
   $H_0,  M: \mathbb{E}(d_{i, uv, m}) = 0 \,,\text{for all} \,u,  v \in M \subset M_0$. 
   \item  Model selection is conducted using the equivalence test $\delta_M$ and the elimination rule $e_M$. Initially, the null hypothesis $H_0: M= M_0$ is proposed. At a significance level $\alpha$, the equivalence test $\delta_M$ is applied to test $H_{0, M}$. If the null hypothesis is not rejected, then $M^*_{1-\alpha} = M$ is defined. Otherwise, the elimination rule $e_M$ is employed to remove rejected models from $M$. This process continues until no further rejection of the null hypothesis occurs, ultimately yielding the surviving models under the MCS procedure.
\end{enumerate}

There are typically two types of test statistics used in the MCS evaluation: the range statistic $T_{R}$ and the semiquadratic statistic $T_{\mathrm{max}}$, which are defined as follows:
$$T_{R}\:=\:\max_{u, v\:\in\:M}\frac{\mid\bar{d}_{i, uv}\mid}{\sqrt{\text{var}(\:d_{i, uv})}}\,,\quad T_{\mathrm{max}}\:=\:\max_{u, v\in M}\frac{\bar{d}_{i, uv}^2}{\text{var}(\begin{array}{c}\bar{d}_{i, uv})\end{array}} $$ 
where $\bar{d}_{i, uv} = M^{-1}\sum_{m=H+1}^{H+M} d_{i, uv, m}$ represents the average loss value between model $u$ and model $v$. The $T_R$ and $T_{\mathrm{max}}$ statistics, along with their corresponding $p$-values, are obtained using the Bootstrap method.

\section{Out-of-sample forecast loss value and MCS test}\label{sec:MCStest}

\begin{table}[ht]
\centering
\caption{Out-of-sample forecast loss value of SSE for the five-step-ahead predictions}
\label{tab:table5}
\resizebox{\textwidth}{!}{
\begin{threeparttable}
\begin{tabular}{lcccccc}
\toprule
Model & MSE & MAE & HMSE & HMAE & QLIKE \\ \midrule
GARCH & $6.535\times 10^{-9}$ & $4.592\times 10^{-5}$ & $1.788$ & $0.909$ & $-8.626$ \\
EGARCH & $5.732\times 10^{-9}$ & $5.463\times 10^{-5}$ & $3.126$ & $1.289$ & $-8.587$ \\
TGARCH & $1.029\times 10^{-8}$ & $5.463\times 10^{-5}$ & $2.692$ & $0.996$ & $-8.543$ \\
GJRGARCH & $1.054\times 10^{-8}$ & $5.387\times 10^{-5}$ & $2.211$ & $0.929$ & $-8.347$ \\

HAR-RV & $7.068\times 10^{-9}$ & $4.904\times 10^{-5}$ & $1.936$ & $1.005$ & $-8.608$ \\
HAR-CJ & $6.220\times 10^{-9}$ & $4.148\times 10^{-5}$ & $1.013$ & $0.746$ & $-8.647$ \\
HAR-RS & $4.453\times 10^{-9}$ & $3.351\times 10^{-5}$ & $0.840$ & $0.580$ & $-8.724$ \\
HAR-REX & $5.113\times 10^{-9}$ & $3.444\times 10^{-5}$ & $0.590$ & $0.554$ & $-8.696$ \\
HAR-REQ & $5.087\times 10^{-9}$ & $3.353\times 10^{-5}$ & $0.527$ & $0.524$ & $-8.697$ \\
HAR-PD-RV & $6.902\times 10^{-9}$ & $3.966\times 10^{-5}$ & $0.805$ & $0.575$ & $-8.606$ \\
HAR-PD-CJ & $5.062\times 10^{-9}$ & $3.257\times 10^{-5}$ & \textbf{\underline{$\mathbf{0.445}$}} & $0.474$ & $-8.699$ \\
HAR-PD-RS & $4.209\times 10^{-9}$ & $3.460\times 10^{-5}$ & $1.039$ & $0.627$ & $-8.659$ \\
HAR-PD-REX & $4.954\times 10^{-9}$ & $3.294\times 10^{-5}$ & $0.525$ & $0.519$ & $-8.700$ \\
HAR-PD-REQ & $4.888\times 10^{-9}$ & \textbf{\underline{$\mathbf{3.127\times 10^{-5}}$}} & $0.454$ & \textbf{\underline{$\mathbf{0.468}$}} & $-8.707$ \\
LASSO-HAR-PD-CJ & $5.072\times 10^{-9}$ & $3.287\times 10^{-5}$ & $0.464$ & $0.484$ & $-8.700$ \\
LASSO-HAR-PD-RS & \textbf{\underline{$\mathbf{4.417\times 10^{-9}}$}} & $3.162\times 10^{-5}$ & $0.673$ & $0.493$ & \textbf{\underline{$\mathbf{-8.732}$}} \\
LASSO-HAR-PD-REX & $5.013\times 10^{-9}$ & $3.325\times 10^{-5}$ & $0.536$ & $0.521$ & $-8.698$ \\
LASSO-HAR-PD-REQ & $5.053\times 10^{-9}$ & $3.473\times 10^{-5}$ & $0.627$ & $0.572$ & $-8.692$ \\
\bottomrule
\end{tabular}
\begin{tablenotes}
\item Note: The bold values in the table indicate the minimum under the corresponding loss function, while the underlined values denote the optimal models identified by the MCS test at a significance level of 0.25.
\end{tablenotes}
\end{threeparttable}
}
\end{table}

\begin{table}[ht]
\centering
\caption{Out-of-sample forecast loss value of SSE for the twenty-two-step-ahead predictions}
\label{tab:table6}
\resizebox{\textwidth}{!}{
\begin{threeparttable}
\begin{tabular}{lcccccc}
\toprule
Model & MSE & MAE & HMSE & HMAE & QLIKE \\ \midrule
GARCH & $7.191 \times 10^{-9}$ & $4.367 \times 10^{-5}$ & $1.349$ & $0.824$ & $-8.619$ \\
EGARCH & $9.048 \times 10^{-9}$ & $6.407 \times 10^{-5}$ & $4.796$ & $1.562$ & $-8.378$ \\
TGARCH & $9.774 \times 10^{-9}$ & $5.295 \times 10^{-5}$ & $0.728$ & $0.706$ & $-7.679$ \\
GJRGARCH & $1.023 \times 10^{-8}$ & $4.689 \times 10^{-5}$ & $0.763$ & $0.607$ & $-8.149$ \\
HAR-RV & $7.057 \times 10^{-9}$ & $4.871 \times 10^{-5}$ & $1.965$ & $1.013$ & $-8.619$ \\
HAR-CJ & $5.082 \times 10^{-9}$ & $3.311 \times 10^{-5}$ & $0.525$ & $0.519$ & $-8.710$ \\
HAR-RS & $4.413 \times 10^{-9}$ & $3.282 \times 10^{-5}$ & $0.846$ & $0.578$ & $-8.736$ \\
HAR-REX & $5.119 \times 10^{-9}$ & $3.396 \times 10^{-5}$ & $0.594$ & $0.554$ & $-8.707$ \\
HAR-REQ & $5.099 \times 10^{-9}$ & $3.308 \times 10^{-5}$ & $0.529$ & $0.524$ & $-8.708$ \\
HAR-PD-RV & $6.986 \times 10^{-9}$ & $3.955 \times 10^{-5}$ & $0.817$ & $0.578$ & $-8.614$ \\
HAR-PD-CJ & $5.062 \times 10^{-9}$ & $3.204 \times 10^{-5}$ & $0.446$ & $0.474$ & $-8.710$ \\
HAR-PD-RS &\textbf{$\mathbf{4.213 \times 10^{-9}}$} & $3.422 \times 10^{-5}$ & $1.059$ & $0.630$ & $-8.669$ \\
HAR-PD-REX & $4.964 \times 10^{-9}$ & $3.253 \times 10^{-5}$ & $0.525$ & $0.510$ & $-8.711$ \\
HAR-PD-REQ & $4.944 \times 10^{-9}$ & $3.102 \times 10^{-5}$ & \textbf{$\underline{\mathbf{0.459}}$} & \textbf{$\underline{\mathbf{0.470}}$} & \textbf{$\underline{\mathbf{-8.717}}$} \\
LASSO-HAR-PD-CJ & $5.068 \times 10^{-9}$ & $3.232 \times 10^{-5}$ & $0.464$ & $0.483$ &${-8.711}$ \\
LASSO-HAR-PD-RS & $\underline{4.377 \times 10^{-9}}$ & \textbf{$\underline{\mathbf{3.097 \times 10^{-5}}}$} & $0.677$ & $0.491$ & $-8.744$ \\
LASSO-HAR-PD-REX & $5.020 \times 10^{-9}$ & $3.277 \times 10^{-5}$ & $0.535$ & $0.519$ & $-8.709$ \\
LASSO-HAR-PD-REQ & $5.081 \times 10^{-9}$ & $3.438 \times 10^{-5}$ & $0.632$ & $0.574$ & $-8.702$\\
\bottomrule
\end{tabular}
\begin{tablenotes}
\item Note: The bold values in the table indicate the minimum under the corresponding loss function, while the underlined values denote the optimal models identified by the MCS test at a significance level of 0.25.
\end{tablenotes}
\end{threeparttable}
}
\end{table}

\begin{table}[ht]
\centering
\caption{MCS test for the one-step-ahead volatility forecasting model of the SSE}
\label{tab:table7}
\resizebox{\textwidth}{!}{
\begin{threeparttable}
\begin{tabular}{@{}llllllllllllll@{}}
\toprule
\multicolumn{1}{c}{\multirow{2}{*}{Model}} & \multicolumn{2}{c}{$\mathrm{QLIKE}_{\alpha_1}$}                       & \multicolumn{2}{c}{$\mathrm{MSE}_{\alpha_1}$}                           & \multicolumn{2}{c}{$\mathrm{QLIKE}_{\alpha_2}$}                         & \multicolumn{2}{c}{$\mathrm{MSE}_{\alpha_2}$}                           & \multicolumn{2}{c}{$\mathrm{QLIKE}_{\alpha_3}$}                         & \multicolumn{2}{c}{$\mathrm{MSE}_{\alpha_3}$}                           \\
\cmidrule(l){2-3} \cmidrule(l){4-5}\cmidrule(l){6-7}\cmidrule(l){8-9}\cmidrule(l){10-11}\cmidrule(l){12-13}
& \multicolumn{1}{c}{$T_{max}$} & \multicolumn{1}{c}{$T_{R}$} & \multicolumn{1}{c}{$T_{max}$} & \multicolumn{1}{c}{$T_{R}$} & \multicolumn{1}{c}{$T_{max}$} & \multicolumn{1}{c}{$T_{R}$} & \multicolumn{1}{c}{$T_{max}$} & \multicolumn{1}{c}{$T_{R}$} & \multicolumn{1}{c}{$T_{max}$} & \multicolumn{1}{c}{$T_{R}$} & \multicolumn{1}{c}{$T_{max}$} & \multicolumn{1}{c}{$T_{R}$} \\
\midrule
GARCH & $0.000$ & $0.000$ & $0.029$ & $0.029$ & $0.034$ & $0.000$ & $0.000$ & $0.000$ & $0.000$ & $0.000$ & $0.000$ & $0.000$ \\
EGARCH & $0.000$ & $0.000$ & $0.067$ & $0.012$ & $0.000$ & $0.000$ & $0.000$ & $0.000$ & $0.000$ & $0.000$ & $0.000$ & $0.000$ \\
TGARCH & $0.000$ & $0.000$ & $0.994$ & $0.090$ & $0.000$ & $0.000$ & $0.000$ & $0.000$ & $0.000$ & $0.000$ & $0.000$ & $0.000$ \\
GJRGARCH & $0.000$ & $0.000$ & $0.822$ & $0.029$ & $0.000$ & $0.000$ & $0.000$ & $0.000$ & $0.000$ & $0.000$ & $0.000$ & $0.000$ \\
HAR-RV & $0.000$ & $0.000$ & $0.209$ & $0.120$ & $0.000$ & $0.000$ & $0.000$ & $0.000$ & $0.000$ & $0.000$ & $0.000$ & $0.000$ \\
HAR-CJ & $0.000$ & $0.000$ & $0.039$ & $0.023$ & $0.000$ & $0.000$ & $0.000$ & $0.000$ & $0.000$ & $0.000$ & $0.000$ & $0.000$ \\
HAR-RS & $1.000$ & $0.004$ & $1.000$ & $0.233$ & $1.000$ & $0.002$ & $1.000$ & $0.181$ & $1.000$ & $0.000$ & $0.893$ & $0.165$ \\
HAR-REX & $0.128$ & $0.308$ & $1.000$ & $0.626$ & $0.000$ & $0.000$ & $1.000$ & $0.482$ & $0.000$ & $0.000$ & $0.918$ & $0.460$ \\
HAR-REQ & $0.486$ & $0.440$ & $1.000$ & $1.000$ & $1.000$ & $0.221$ & $1.000$ & $1.000$ & $0.000$ & $0.000$ & $1.000$ & $1.000$ \\
HAR-PD-RV & $0.000$ & $0.000$ & $1.000$ & $0.233$ & $0.000$ & $0.000$ & $0.107$ & $0.098$ & $0.000$ & $0.000$ & $0.000$ & $0.000$ \\
HAR-PD-CJ & $1.000$ & $1.000$ & $1.000$ & $1.000$ & $0.000$ & $1.000$ & $0.735$ & $1.000$ & $1.000$ & $0.992$ & $0.499$ & $1.000$ \\
HAR-PD-RS & $0.694$ & $0.004$ & $1.000$ & $1.000$ & $0.427$ & $0.437$ & $1.000$ & $1.000$ & $0.340$ & $0.399$ & $1.000$ & $1.000$ \\
HAR-PD-REX & $0.994$ & $1.000$ & $1.000$ & $\mathbf{1.000}$ & $0.802$ & $0.656$ & $\mathbf{1.000}$ & $\mathbf{1.000}$ & $0.678$ & $0.607$ & $\mathbf{1.000}$ & $\mathbf{1.000}$ \\
HAR-PD-REQ & $1.000$ & $1.000$ & $1.000$ & $1.000$ & $1.000$ & $1.000$ & $1.000$ & $1.000$ & $1.000$ & $1.000$ & $1.000$ & $1.000$ \\
LASSO-HAR-PD-CJ & $0.672$ & $0.660$ & $1.000$ & $1.000$ & $0.380$ & $0.547$ & $1.000$ & $0.975$ & $0.329$ & $0.399$ & $0.998$ & $0.977$ \\
LASSO-HAR-PD-RS & $\mathbf{1.000}$ & $\mathbf{1.000}$ & $\mathbf{1.000}$ & $1.000$ & $\mathbf{1.000}$ & $\mathbf{1.000}$ & $1.000$ & $1.000$ & $\mathbf{1.000}$ & $\mathbf{1.000}$ & $0.997$ & $1.000$ \\
LASSO-HAR-PD-REX & $0.892$ & $0.660$ & $1.000$ & $1.000$ & $0.000$ & $0.000$ & $1.000$ & $1.000$ & $0.333$ & $0.399$ & $1.000$ & $1.000$ \\
LASSO-HAR-PD-REQ & $0.046$ & $0.004$ & $1.000$ & $1.000$ & $0.432$ & $0.449$ & $1.000$ & $1.000$ & $0.000$ & $0.000$ & $0.807$ & $1.000$ \\
\bottomrule
\end{tabular}
\begin{tablenotes} 
   \item{Note: $\alpha_1$, $\alpha_2$ and $\alpha_3$ denote that the significance level of $\mathrm{MCS}$ test is taken as $0. 01$, $0. 1$ and $0. 25$, respectively. The $p$ value indicates the prediction accuracy of the model, and is retained to three decimal places; the higher the $p$ value, the better the prediction accuracy. The optimal model under the test is bolded to indicate that the model has the optimal volatility prediction performance in the corresponding prediction interval.}
\end{tablenotes}

\end{threeparttable}
}
\end{table}

\begin{table}[ht]
\centering
\caption{MCS test for the five-step-ahead volatility forecasting model of the SSE}
\label{tab:table8}
\resizebox{\textwidth}{!}{
\begin{threeparttable}
\begin{tabular}{@{}llllllllllllll@{}}
\toprule
\multicolumn{1}{c}{\multirow{2}{*}{Model}} & \multicolumn{2}{c}{$\mathrm{QLIKE}_{\alpha_1}$}                       & \multicolumn{2}{c}{$\mathrm{MSE}_{\alpha_1}$}                           & \multicolumn{2}{c}{$\mathrm{QLIKE}_{\alpha_2}$}                         & \multicolumn{2}{c}{$\mathrm{MSE}_{\alpha_2}$}                           & \multicolumn{2}{c}{$\mathrm{QLIKE}_{\alpha_3}$}                         & \multicolumn{2}{c}{$\mathrm{MSE}_{\alpha_3}$}                           \\
\cmidrule(l){2-3} \cmidrule(l){4-5}\cmidrule(l){6-7}\cmidrule(l){8-9}\cmidrule(l){10-11}\cmidrule(l){12-13}
& \multicolumn{1}{c}{$T_{max}$} & \multicolumn{1}{c}{$T_{R}$} & \multicolumn{1}{c}{$T_{max}$} & \multicolumn{1}{c}{$T_{R}$} & \multicolumn{1}{c}{$T_{max}$} & \multicolumn{1}{c}{$T_{R}$} & \multicolumn{1}{c}{$T_{max}$} & \multicolumn{1}{c}{$T_{R}$} & \multicolumn{1}{c}{$T_{max}$} & \multicolumn{1}{c}{$T_{R}$} & \multicolumn{1}{c}{$T_{max}$} & \multicolumn{1}{c}{$T_{R}$} \\
\midrule
GARCH & $0.000$ & $0.000$ & $0.000$ & $0.000$ & $0.034$ & $0.029$ & $0.000$ & $0.000$ & $0.000$ & $0.000$ & $0.016$ & $0.021$ \\
EGARCH & $0.000$ & $0.000$ & $0.099$ & $0.013$ & $0.000$ & $0.000$ & $0.000$ & $0.000$ & $0.000$ & $0.000$ & $0.000$ & $0.000$ \\
TGARCH & $0.000$ & $0.000$ & $0.046$ & $0.072$ & $0.000$ & $0.000$ & $0.000$ & $0.000$ & $0.000$ & $0.000$ & $0.000$ & $0.000$ \\
GJRGARCH & $0.000$ & $0.000$ & $0.964$ & $0.029$ & $0.000$ & $0.000$ & $0.000$ & $0.000$ & $0.000$ & $0.000$ & $0.043$ & $0.083$ \\

HAR-RV & $0.000$ & $0.000$ & $0.298$ & $0.080$ & $0.000$ & $0.000$ & $0.000$ & $0.000$ & $0.000$ & $0.000$ & $0.000$ & $0.000$ \\
HAR-CJ & $0.251$ & $0.994$ & $1.000$ & $0.469$ & $0.000$ & $0.000$ & $1.000$ & $0.423$ & $0.192$ & $0.659$ & $0.993$ & $0.486$ \\
HAR-RS & $1.000$ & $0.008$ & $1.000$ & $0.098$ & $1.000$ & $0.000$ & $1.000$ & $0.085$ & $1.000$ & $0.002$ & $0.998$ & $0.096$ \\
HAR-REX & $0.129$ & $0.647$ & $1.000$ & $0.530$ & $0.000$ & $0.000$ & $1.000$ & $0.423$ & $0.084$ & $0.659$ & $1.000$ & $0.602$ \\
HAR-REQ & $0.664$ & $1.000$ & $1.000$ & $1.000$ & $0.237$ & $0.640$ & $1.000$ & $1.000$ & $0.403$ & $0.779$ & $1.000$ & $0.931$ \\
HAR-PD-RV & $0.000$ & $0.000$ & $0.246$ & $0.469$ & $0.000$ & $0.000$ & $0.107$ & $0.097$ & $0.000$ & $0.000$ & $0.117$ & $0.118$ \\
HAR-PD-CJ & $0.882$ & $1.000$ & $1.000$ & $1.000$ & $1.000$ & $1.000$ & $0.802$ & $1.000$ & $0.717$ & $0.805$ & $1.000$ & $0.724$ \\
HAR-PD-RS & $0.943$ & $0.005$ & $1.000$ & $1.000$ & $0.665$ & $0.640$ & $1.000$ & $1.000$ & $0.989$ & $0.001$ & $1.000$ & $0.997$ \\
HAR-PD-REX & $1.000$ & $1.000$ & $1.000$ & $\textbf{1.000}$ & $1.000$ & $\textbf{1.000}$ & $1.000$ & $\textbf{1.000}$ & $0.995$ & $0.998$ & $\textbf{1.000}$ & $\textbf{1.000}$ \\
HAR-PD-REQ & $1.000$ & $1.000$ & $\textbf{1.000}$ & $1.000$ & $1.000$ & $1.000$ & $1.000$ & $1.000$ & $1.000$ & $1.000$ & $1.000$ & $0.989$ \\
LASSO-HAR-PD-CJ & $0.794$ & $1.000$ & $1.000$ & $1.000$ & $0.345$ & $0.995$ & $0.971$ & $0.440$ & $0.654$ & $0.781$ & $0.995$ & $0.646$ \\
LASSO-HAR-PD-RS & $\textbf{1.000}$ & $\textbf{1.000}$ & $1.000$ & $1.000$ & $\textbf{1.000}$ & $1.000$ & $1.000$ & $1.000$ & $\textbf{1.000}$ & $\textbf{1.000}$ & $0.993$ & $1.000$ \\
LASSO-HAR-PD-REX & $0.994$ & $1.000$ & $1.000$ & $1.000$ & $0.644$ & $0.640$ & $\textbf{1.000}$ & $1.000$ & $0.987$ & $0.779$ & $1.000$ & $0.971$ \\
LASSO-HAR-PD-REQ & $0.062$ & $0.005$ & $1.000$ & $1.000$ & $0.000$ & $0.000$ & $1.000$ & $1.000$ & $0.044$ & $0.889$ & $0.988$ & $0.592$ \\
\bottomrule
\end{tabular}
\begin{tablenotes} 
   \item{Note: $\alpha_1$, $\alpha_2$ and $\alpha_3$ denote that the significance level of $\mathrm{MCS}$ test is taken as $0. 01$, $0. 1$ and $0. 25$, respectively. The $p$ value indicates the prediction accuracy of the model, and is retained to three decimal places; the higher the $p$ value, the better the prediction accuracy. The optimal model under the test is bolded to indicate that the model has the optimal volatility prediction performance in the corresponding prediction interval.}
\end{tablenotes}

\end{threeparttable}
}
\end{table}

\begin{table}[ht]
\centering
\caption{MCS test for the twenty-two-step-ahead volatility forecasting model of the SSE}
\label{tab:table9}
\resizebox{\textwidth}{!}{
\begin{threeparttable}
\begin{tabular}{@{}llllllllllllll@{}}
\toprule
\multicolumn{1}{c}{\multirow{2}{*}{Model}} & \multicolumn{2}{c}{$\mathrm{QLIKE}_{\alpha_1}$}                       & \multicolumn{2}{c}{$\mathrm{MSE}_{\alpha_1}$}                           & \multicolumn{2}{c}{$\mathrm{QLIKE}_{\alpha_2}$}                         & \multicolumn{2}{c}{$\mathrm{MSE}_{\alpha_2}$}                           & \multicolumn{2}{c}{$\mathrm{QLIKE}_{\alpha_3}$}                         & \multicolumn{2}{c}{$\mathrm{MSE}_{\alpha_3}$}                           \\
\cmidrule(l){2-3} \cmidrule(l){4-5}\cmidrule(l){6-7}\cmidrule(l){8-9}\cmidrule(l){10-11}\cmidrule(l){12-13}
& \multicolumn{1}{c}{$T_{max}$} & \multicolumn{1}{c}{$T_{R}$} & \multicolumn{1}{c}{$T_{max}$} & \multicolumn{1}{c}{$T_{R}$} & \multicolumn{1}{c}{$T_{max}$} & \multicolumn{1}{c}{$T_{R}$} & \multicolumn{1}{c}{$T_{max}$} & \multicolumn{1}{c}{$T_{R}$} & \multicolumn{1}{c}{$T_{max}$} & \multicolumn{1}{c}{$T_{R}$} & \multicolumn{1}{c}{$T_{max}$} & \multicolumn{1}{c}{$T_{R}$} \\
\midrule
GARCH & $0.019$ & $0.045$ & $0.000$ & $0.000$ & $0.034$ & $0.029$ & $0.000$ & $0.000$ & $0.000$ & $0.000$ & $0.000$ & $0.000$ \\
EGARCH & $0.000$ & $0.000$ & $0.099$ & $0.013$ & $0.000$ & $0.000$ & $0.000$ & $0.000$ & $0.000$ & $0.000$ & $0.000$ & $0.000$ \\
TGARCH & $0.113$ & $0.034$ & $0.047$ & $0.067$ & $0.000$ & $0.000$ & $0.000$ & $0.000$ & $0.000$ & $0.000$ & $0.000$ & $0.000$ \\
GJRGARCH & $0.000$ & $0.000$ & $0.964$ & $0.029$ & $0.000$ & $0.000$ & $0.000$ & $0.000$ & $0.000$ & $0.000$ & $0.000$ & $0.000$ \\

HAR-RV & $0.122$ & $0.125$ & $0.278$ & $0.079$ & $0.000$ & $0.000$ & $0.000$ & $0.000$ & $0.000$ & $0.000$ & $0.000$ & $0.000$ \\
HAR-CJ & $1.000$ & $0.660$ & $1.000$ & $0.453$ & $0.000$ & $0.000$ & $0.990$ & $0.559$ & $0.000$ & $0.000$ & $0.993$ & $0.559$ \\
HAR-RS & $1.000$ & $0.415$ & $1.000$ & $0.095$ & $1.000$ & $0.002$ & $0.972$ & $0.359$ & $0.000$ & $0.000$ & $0.972$ & $0.359$ \\
HAR-REX & $1.000$ & $0.573$ & $1.000$ & $0.946$ & $0.000$ & $0.000$ & $0.816$ & $0.481$ & $0.000$ & $0.000$ & $0.816$ & $0.481$ \\
HAR-REQ & $1.000$ & $0.937$ & $0.246$ & $0.453$ & $0.123$ & $0.513$ & $1.000$ & $0.878$ & $0.000$ & $0.000$ & $1.000$ & $0.877$ \\
HAR-PD-RV & $0.971$ & $0.124$ & $1.000$ & $0.485$ & $0.000$ & $0.000$ & $0.000$ & $0.000$ & $0.000$ & $0.000$ & $0.000$ & $0.000$ \\
HAR-PD-CJ & $1.000$ & $0.750$ & $1.000$ & $1.000$ & $0.303$ & $0.548$ & $1.000$ & $0.672$ & $0.000$ & $0.000$ & $1.000$ & $0.652$ \\
HAR-PD-RS & $1.000$ & $0.998$ & $1.000$ & $1.000$ & $0.797$ & $0.003$ & $1.000$ & $0.997$ & $0.000$ & $0.000$ & $1.000$ & $0.997$ \\
HAR-PD-REX & $1.000$ & $\mathbf{1.000}$ & $1.000$ & $0.914$ & $0.995$ & $\mathbf{1.000}$ & $\mathbf{1.000}$ & $\mathbf{1.000}$ & $0.000$ & $0.000$ & $\mathbf{1.000}$ & $\mathbf{1.000}$ \\
HAR-PD-REQ & $\mathbf{1.000}$ & $0.995$ & $\mathbf{1.000}$ & $\mathbf{1.000}$ & $1.000$ & $1.000$ & $1.000$ & $0.985$ & $0.000$ & $0.000$ & $1.000$ & $0.985$ \\
LASSO-HAR-PD-CJ & $0.000$ & $0.000$ & $1.000$ & $1.000$ & $0.259$ & $0.513$ & $0.996$ & $0.615$ & $0.000$ & $0.000$ & $0.995$ & $0.978$ \\
LASSO-HAR-PD-RS & $1.000$ & $0.990$ & $1.000$ & $1.000$ & $\mathbf{1.000}$ & $1.000$ & $1.000$ & $0.978$ & $\mathbf{1.000}$ & $\mathbf{1.000}$ & $0.993$ & $0.978$ \\
LASSO-HAR-PD-REX & $1.000$ & $0.987$ & $1.000$ & $1.000$ & $0.604$ & $0.548$ & $1.000$ & $1.000$ & $0.000$ & $0.000$ & $1.000$ & $0.971$ \\
LASSO-HAR-PD-REQ & $0.000$ & $0.000$ & $1.000$ & $1.000$ & $0.000$ & $0.000$ & $0.612$ & $0.621$ & $0.000$ & $0.000$ & $0.612$ & $0.621$ \\
\bottomrule
\end{tabular}
\begin{tablenotes} 
 \item{Note: $\alpha_1$, $\alpha_2$ and $\alpha_3$ denote that the significance level of $\mathrm{MCS}$ test is taken as $0. 01$, $0. 1$ and $0. 25$, respectively. The $p$ value indicates the prediction accuracy of the model, and is retained to three decimal places; the higher the $p$ value, the better the prediction accuracy. The optimal model under the test is bolded to indicate that the model has the optimal volatility prediction performance in the corresponding prediction interval. }
\end{tablenotes}
\end{threeparttable}
}
\end{table}

\begin{table}[ht]
\centering
\caption{Out-of-sample forecast loss value of CSI 300 for the five-step-ahead predictions}
\label{tab:table13}
\resizebox{\textwidth}{!}{
\begin{threeparttable}
\begin{tabular}{lccccc}
\toprule
Model & MSE & MAE & HMSE & HMAE & QLIKE \\ 
\midrule
GARCH & $1.176 \times 10^{-8}$ & $5.154 \times 10^{-5}$ & $1.584$ & $0.746$ & $-8.392$ \\
EGARCH & $1.651 \times 10^{-8}$ & $6.938 \times 10^{-5}$ & $3.995$ & $1.209$ & $-8.339$ \\
TGARCH & $1.027 \times 10^{-8}$ & $4.512 \times 10^{-5}$ & $0.892$ & $0.647$ & $-8.483$ \\
GJRGARCH & $1.133\times 10^{-8}$ & $4.905\times 10^{-5}$ &$1.247$ & $0.673$ &$-8.403$\\
HAR-RV & $1.286 \times 10^{-8}$ & $5.516 \times 10^{-5}$ & $3.669$ & $0.878$ & $-8.367$ \\
HAR-CJ & $9.018 \times 10^{-9}$ & $3.846 \times 10^{-5}$ & $0.476$ & $0.497$ & $-8.543$ \\
HAR-RS & $7.581 \times 10^{-9}$ & \textbf{$\mathbf{3.795 \times 10^{-5}}$} & $0.832$ & $0.552$ & $-8.514$ \\
HAR-REX & $9.645 \times 10^{-9}$ & $4.461 \times 10^{-5}$ & $0.807$ & $0.671$ & $-8.519$ \\
HAR-REQ & $9.352 \times 10^{-9}$ & $\underline{3.840 \times 10^{-5}}$ & \textbf{$\underline{\mathbf{0.433}}$} & \textbf{$\underline{\mathbf{0.485}}$} & $-8.531$ \\
HAR-PD-RV & $9.985 \times 10^{-9}$ & $4.602 \times 10^{-5}$ & $1.257$ & $0.736$ & $-8.507$ \\
HAR-PD-CJ & $8.873 \times 10^{-9}$ & $4.258 \times 10^{-5}$ & $0.696$ & $0.611$ & $-8.533$ \\
HAR-PD-RS & \textbf{$\underline{\mathbf{7.575 \times 10^{-9}}}$} & $3.907 \times 10^{-5}$ & $0.948$ & $0.583$ & $-8.541$ \\
HAR-PD-REX & $8.502 \times 10^{-9}$ & $4.152 \times 10^{-5}$ & $0.662$ & $0.607$ & $-8.540$ \\
HAR-PD-REQ & $8.179 \times 10^{-9}$ & $4.040 \times 10^{-5}$ & $0.632$ & $0.585$ &\textbf{ $\underline{\mathbf{-8.559}}$} \\
LASSO-HAR-PD-CJ & $8.878 \times 10^{-9}$ & $4.282 \times 10^{-5}$ & $0.712$ & $0.619$ & $-8.533$ \\
LASSO-HAR-PD-RS & $7.576 \times 10^{-9}$ & $3.907 \times 10^{-5}$ & $0.948$ & $0.581$ & $-8.542$ \\
LASSO-HAR-PD-REX & $8.745 \times 10^{-9}$ & $4.247 \times 10^{-5}$ & $0.682$ & $0.618$ & $-8.531$ \\
LASSO-HAR-PD-REQ & $8.182 \times 10^{-9}$ & $4.034 \times 10^{-5}$ & $0.629$ & $0.583$ & $-8.558$ \\
\bottomrule
\end{tabular}
\begin{tablenotes}
\item Note: The bold values in the table indicate the minimum under the corresponding loss function, while the underlined values denote the optimal models identified by the MCS test at a significance level of 0.25.
\end{tablenotes}
\end{threeparttable}
}
\end{table}

\begin{table}[!ht]
\centering
\caption{Out-of-sample forecast loss value of CSI 300 for the twenty-two-step-ahead predictions}
\label{tab:table14}
\resizebox{\textwidth}{!}{
\begin{threeparttable}
\begin{tabular}{lccccc}
\toprule
Model & MSE & MAE & HMSE & HMAE & QLIKE \\\midrule
GARCH & $1.128 \times 10^{-8}$ & $4.319 \times 10^{-5}$ & $0.539$ & $0.501$ & $-8.301$ \\ 
EGARCH & $1.578 \times 10^{-8}$ & $7.779\times10^{-5}$ & $4.720 $ & $1.440$ & $-8.115$ \\ 
TGARCH & $1.071 \times 10^{-8}$ & $4.048\times 10^{-5}$ & $0.656$ & $0.589$ & $-8.178$ \\ 
GJRGARCH & $1.138 \times 10^{-8}$ & $4.336 \times 10^{-5}$ & $0.506 $& $0.493$ &$-8.269$  \\
HAR-RV & $1.264 \times 10^{-8}$ & $5.235 \times 10^{-5}$ & $3.818 $ & $0.886$ & $-8.417$ \\ 
HAR-CJ & $9.098 \times 10^{-9}$ & $3.647 \times 10^{-5}$ & $0.483$ & $0.498$ & $-8.590$ \\ 
HAR-RS & $7.613 \times 10^{-9}$ & $3.638 \times 10^{-5}$ & $0.814$ & $0.553$ & $-8.575$ \\ 
HAR-REX & $9.566 \times 10^{-9}$ & $4.221 \times 10^{-5}$ & $0.800$ & $0.668$ & $-8.567$ \\ 
HAR-REQ & $9.380 \times 10^{-9}$ & \textbf{$\underline{\mathbf{3.599 \times 10^{-5}}}$} & \textbf{$\underline{\mathbf{0.434}}$} & \textbf{$\underline{\mathbf{0.482}}$} & $-8.583$ \\ 
HAR-PD-RV & $9.664 \times 10^{-9}$ & $4.243 \times 10^{-5}$ & $1.269$ & $0.732$ & $-8.557$ \\ 
HAR-PD-CJ & $9.071 \times 10^{-9}$ & $4.154 \times 10^{-5}$ & $0.766$ & $0.635$ & $-8.577$ \\ 
HAR-PD-RS &\textbf{$\underline{\mathbf{7.625 \times 10^{-9}}}$} & $3.705\times 10^{-5}$ & $0.941$ & $0.574$ & $-8.599$\\
HAR-PD-REX & $8.751 \times 10^{-9}$ & $4.087 \times 10^{-5}$ & $0.716$ & $0.629$ & $-8.592$ \\ 
HAR-PD-REQ & $8.536 \times 10^{-9}$ & $4.077 \times 10^{-5}$ & $0.752$ & $0.627$ & \textbf{$\underline{\mathbf{-8.601}}$} \\ 
LASSO-HAR-PD-CJ & $9.075 \times 10^{-9}$ & $4.177 \times 10^{-5}$ & $0.782$ & $0.643$ & $-8.577$ \\ 
LASSO-HAR-PD-RS & $8.176 \times 10^{-9}$ & $3.687 \times 10^{-5}$ & $0.724$ & $0.559$ & $-8.597$ \\
LASSO-HAR-PD-REX & $9.124 \times 10^{-9}$ & $4.334 \times 10^{-5}$ & $0.790$ & $0.667$ & $-8.570$ \\ 
LASSO-HAR-PD-REQ& $8.556 \times 10^{-9}$ & $4.080 \times 10^{-5}$ & $0.750$ & $0.625$ & $-8.600$ \\ 

\bottomrule
\end{tabular}
\begin{tablenotes}
\item Note: The bold values in the table indicate the minimum under the corresponding loss function, while the underlined values denote the optimal models identified by the MCS test at a significance level of 0.25.
\end{tablenotes}
\end{threeparttable}
}
\end{table}

\begin{table}[ht]
\centering
\caption{MCS test for each volatility model of CSI 300 with a out-of-sample of 300-day}
\label{tab:table15}
\resizebox{\textwidth}{!}{
\begin{threeparttable}
\begin{tabular}{@{}llllllllllllll@{}}
\toprule
\multicolumn{1}{c}{\multirow{3}{*}{Model}} & \multicolumn{4}{c}{One-step-ahead} & \multicolumn{4}{c}{Five-step-ahead} & \multicolumn{4}{c}{Twenty-two-step-ahead} \\
\cmidrule(lr){2-5} \cmidrule(lr){6-9} \cmidrule(lr){10-13}
& \multicolumn{2}{c}{$\mathrm{QLIKE}$} & \multicolumn{2}{c}{$\mathrm{MSE}$} & \multicolumn{2}{c}{$\mathrm{QLIKE}$} & \multicolumn{2}{c}{$\mathrm{MSE}$} & \multicolumn{2}{c}{$\mathrm{QLIKE}$} & \multicolumn{2}{c}{$\mathrm{MSE}$} \\
\cmidrule(l){2-3} \cmidrule(l){4-5} \cmidrule(l){6-7} \cmidrule(l){8-9} \cmidrule(l){10-11} \cmidrule(l){12-13}
& \multicolumn{1}{c}{$T_{max}$} & \multicolumn{1}{c}{$T_{R}$} & \multicolumn{1}{c}{$T_{max}$} & \multicolumn{1}{c}{$T_{R}$} & \multicolumn{1}{c}{$T_{max}$} & \multicolumn{1}{c}{$T_{R}$} & \multicolumn{1}{c}{$T_{max}$} & \multicolumn{1}{c}{$T_{R}$} & \multicolumn{1}{c}{$T_{max}$} & \multicolumn{1}{c}{$T_{R}$} & \multicolumn{1}{c}{$T_{max}$} & \multicolumn{1}{c}{$T_{R}$} \\
\midrule
GARCH & $0.000$ & $0.000$ & $0.000$ & $0.000$ & $0.034$ & $0.029$ & $0.000$ & $0.000$ & $0.000$ & $0.000$ & $0.000$ & $0.000$ \\
EGARCH & $0.000$ & $0.000$ & $0.121$ & $0.214$ & $0.000$ & $0.000$ & $0.000$ & $0.000$ & $0.000$ & $0.000$ & $0.000$ & $0.000$ \\
TGARCH & $0.000$ & $0.000$ & $0.047$ & $0.067$ & $0.000$ & $0.000$ & $0.000$ & $0.000$ & $0.000$ & $0.000$ & $0.841$ & $0.558$ \\
GJRGARCH & $0.000$ & $0.000$ & $0.964$ & $0.029$ & $0.000$ & $0.000$ & $0.000$ & $0.000$ & $0.000$ & $0.000$ & $0.000$ & $0.000$ \\

HAR-RV & $0.000$ & $0.000$ & $0.278$ & $0.079$ & $0.000$ & $0.000$ & $0.000$ & $0.000$ & $0.000$ & $0.000$ & $0.000$ & $0.000$ \\
HAR-CJ & $1.000$ & $0.986$ & $1.000$ & $0.982$ & $1.000$ & $1.000$ & $0.990$ & $0.559$ & $0.992$ & $0.998$ & $0.993$ & $0.999$ \\
HAR-RS & $1.000$ & $0.998$ & $0.987$ & $1.000$ & $1.000$ & $1.000$ & $0.972$ & $0.359$ & $1.000$ & $0.998$ & $0.972$ & $1.000$ \\
HAR-REX & $0.000$ & $0.000$ & $0.934$ & $0.458$ & $0.791$ & $0.427$ & $0.816$ & $0.481$ & $0.984$ & $0.432$ & $0.816$ & $0.481$ \\
HAR-REQ & $1.000$ & $0.997$ & $1.000$ & $0.982$ & $1.000$ & $1.000$ & $1.000$ & $0.878$ & $\mathbf{1.000}$ & $\mathbf{1.000}$ & $1.000$ & $0.877$ \\
HAR-PD-RV & $0.000$ & $0.000$ & $0.819$ & $0.911$ & $0.000$ & $0.673$ & $0.640$ & $0.000$ & $0.993$ & $0.998$ & $0.851$ & $0.000$ \\
HAR-PD-CJ & $0.365$ & $0.546$ & $1.000$ & $0.755$ & $0.303$ & $0.894$ & $0.848$ & $0.672$ & $0.963$ & $0.993$ & $0.611$ & $0.417$ \\
HAR-PD-RS & $\mathbf{1.000}$ & $\mathbf{1.000}$ & $0.981$ & $0.982$ & $1.000$ & $0.979$ & $1.000$ & $0.997$ & $1.000$ & $1.000$ & $1.000$ & $0.997$ \\
HAR-PD-REX & $0.796$ & $0.816$ & $1.000$ & $0.985$ & $0.995$ & $\mathbf{1.000}$ & $\mathbf{1.000}$ & $\mathbf{1.000}$ & $1.000$ & $1.000$ & ${0.985}$ & ${0.871}$ \\
HAR-PD-REQ & $1.000$ & $1.000$ & $\mathbf{1.000}$ & $1.000$ & $1.000$ & $1.000$ & $1.000$ & $0.985$ & $0.000$ & $0.000$ & $1.000$ & $0.985$ \\
LASSO-HAR-PD-CJ & $0.296$ & $0.505$ & $\mathbf{1.000}$ & $0.191$ & $0.829$ & $0.811$ & $0.996$ & $0.615$ & $0.937$ & $0.998$ & $0.995$ & $0.978$ \\
LASSO-HAR-PD-RS & $\mathbf{1.000}$ & $\mathbf{1.000}$ & $0.927$ & $0.826$ & $\mathbf{1.000}$ & $1.000$ & $1.000$ & $0.978$ & ${1.000}$ & ${1.000}$ & $\mathbf{1.000}$ & $\mathbf{1.000}$ \\
LASSO-HAR-PD-REX & $0.366$ & $0.504$ & $1.000$ & $0.755$ & $0.715$ & $0.640$ & $1.000$ & $1.000$ & $0.499$ & $0.411$ & $1.000$ & $0.971$ \\
LASSO-HAR-PD-REQ & $1.000$ & $0.927$ & $\mathbf{1.000}$ & $\mathbf{1.000}$ & $1.000$ & $0.979$ & $0.612$ & $0.621$ & $1.000$ & $0.998$ & $0.612$ & $0.621$ \\
\bottomrule
\end{tabular}
\begin{tablenotes} 
   \item{Note: The above are the results of MCS test at the confidence level of $0.25$. The $p$-value indicates the prediction accuracy of the model, and is retained to three decimal places; the higher the $p$-value, the better the prediction accuracy. The largest $p$ value is marked in bold to indicate that the model has the optimal volatility prediction performance in the corresponding prediction interval. }
\end{tablenotes}
\end{threeparttable}
}
\end{table}

\end{document}